%% file: main.tex
\documentclass[sigconf]{acmart}

\usepackage{subcaption}

\AtBeginDocument{%
  \providecommand\BibTeX{{%
    \normalfont B\kern-0.5em{\scshape i\kern-0.25em b}\kern-0.8em\TeX}}}

\setcopyright{acmcopyright}
\copyrightyear{2023}
\acmYear{2023}
\acmDOI{XXXXXXX.XXXXXXX}

\acmConference[AIES '23]{AAAI/ACM Conference on Artificial Intelligence, Ethics, and Society}{August, 2023}{Montreal, QC, Canada}
\acmBooktitle{Proceedings of the AAI/ACM Conference on Artificial Intelligence, Ethics, and Society, August 08--10, 2023, Montreal, QC} 
\acmPrice{15.00}
\acmISBN{978-1-4503-XXXX-X/18/06}

\begin{document}

\title{Machine Learning practices and infrastructures}

\author{Glen Berman}
\email{glen.berman@anu.edu.au}
\orcid{0003-3249-0190}
\affiliation{%
  \institution{Australian National University}
  \city{Canberra}
  \state{ACT}
  \country{Australia}
}

\renewcommand{\shortauthors}{Berman}

\begin{abstract}
Machine Learning (ML) systems, particularly when deployed in high-stakes domains, are deeply consequential. They can exacerbate existing inequities, create new modes of discrimination, and reify outdated social constructs. Accordingly, the social context (i.e. organisations, teams, cultures) in which ML systems are developed is a site of active research for the field of AI ethics, and intervention for policymakers. This paper focuses on one aspect of social context that is often overlooked: interactions between practitioners and the tools they rely on, and the role these interactions play in shaping ML practices and the development of ML systems. In particular, through an empirical study of questions asked on the Stack Exchange forums, the use of interactive computing platforms (e.g. Jupyter Notebook and Google Colab) in ML practices is explored. I find that interactive computing platforms are used in a host of learning and coordination practices, which constitutes an infrastructural relationship between interactive computing platforms and ML practitioners. I describe how ML practices are co-evolving alongside the development of interactive computing platforms, and highlight how this risks making invisible aspects of the ML life cycle that AI ethics researchers' have demonstrated to be particularly salient for the societal impact of deployed ML systems.\looseness=-1

\end{abstract}



\maketitle

\section{Introduction}
\label{sec:introduction}

\input{sections/1_intro}

\section{Related work}

\label{sec:related_work}
\input{sections/2_related_work.tex}

\section{Studying infrastructures \& practices}

\label{sec:theory}
\input{sections/3_theoretical_lens}

\section{Method}

\label{sec:method}
\input{sections/4_method}

\section{Findings}

\label{sec:findings}
\input{sections/5_findings}

\section{Discussion}

\label{sec:discussion}
\input{sections/6_discussion}

\section{Limitations}

\label{sec:limitations}
\input{sections/7_limitations}

\section{Conclusion}

\label{sec:conclusion}
\input{sections/8_conclusion}

\begin{acks}
This research is part of a larger PhD research project, supported by the Australian Government Research Training Program Scholarship. I acknowledge feedback generously provided by Jochen Trumpf, Jenny Davis, Ben Hutchinson, Kate Williams, Charlotte Bradley, Ned Cooper, Kathy Reid, and the anonymous reviewers.
\end{acks}

\bibliographystyle{ACM-Reference-Format}
\bibliography{references}

\appendix

\label{sec:appendix}

\input{sections/9_appendix}

\end{document}

%% file: sections/1_intro.tex
It follows from the notion that Machine Learning (ML) systems ought to be thought of as \textit{sociotechnical} systems \cite{selbst2019fairness}---i.e. systems that are socially constructed, requiring both human actors and machines to work \cite{emery1993characteristics}---that the social context in which an ML system is researched, developed, and deployed is likely to shape the characteristics of that system. Given the increasing rate of ML system deployment in high-stakes domains, and widespread evidence of ML systems failing to meet societal expectations \cite[e.g.][]{buolamwini2018gender, obermeyer2019dissecting, koenecke2020racial, shelby2022sociotechnical}, a key question for the ML field relates to infrastructuralisation and its implications for ML practices and deployed ML systems. This paper begins to address this question by attending to one aspect of social context---interactions between ML practitioners and the tools they use to research, build, and deploy ML systems---and demonstrating the relevance of this context to concerns raised by AI ethics researchers.

The social context of ML system development has been studied in the emerging AI ethics field \cite[e.g.][]{hopkins2021machine, li2021algorithmic, bessen2022cost, deshpande2022responsible}. However, relatively little attention has been paid to tracing the relationship between specific material features of this context and the characteristics of ML systems that are developed \cite{langenkamp2022how}. That is, the role of material things (e.g. software tools, office layouts, computer interfaces, network connections), which themselves are socially constructed, alongside social things (e.g. people, beliefs, norms) in shaping ML systems merits closer scrutiny. In this paper, I consider one aspect of this sociomaterial context of ML system development: the use of interactive computing platforms (e.g. Jupyter Notebooks and Google Colaboratory) during ML model development and evaluation. I explore the structure of these platforms and their use by ML practitioners, and consider the ways in which this use may contribute to conventions of ML practices. This exploration serves to illustrate the importance for the AI ethics field of attending both to the sociomaterial context of ML system development generally, and to the role of interactive computing platforms, in particular. The research question to which this exploration is addressed is: \textit{how are interactive computing platforms used in ML practices?}

To answer this question I developed a probabilistic topic model of user-contributed questions on the Stack Exchange forums related to ML and the use of interactive computing platforms. Stack Exchange forums were selected due to their wide use by data and computer scientists, software engineers, and technologists generally \cite{anderson2012discovering, barua2014what}. Alongside this I undertook qualitative text analysis of a small sample of Stack Exchange questions. I find that interactive computing platforms are used in a range of ML practices, particularly in the data curation and processing, and model training and evaluation stages of ML system development. I highlight the role of interactive computing platforms in learning practices, and in practices of coordination across multiple infrastructures. To interpret these findings I draw on sociological studies of infrastructures and practices, particularly the work of sociologists Susan Leigh Star \cite{star1989structure, star1996steps, star1999ethnography, bowker2000sorting} and Elizabeth Shove \cite{shove2003comfort, shove2016matters, watson2022how}, and cultural anthropologist Brian Larkin \cite{larkin2013politics, larkin2020promising}. I conclude that learning and coordination roles are indicative of an infrastructural relationship between ML practitioners and interactive computing platforms, which renders some of the aspects of ML systems development that AI ethics discourse has highlighted as particularly consequential (e.g. the importance of training dataset provenance \cite{gebru2021datasheets, denton2021genealogy}) as invisible to ML practitioners. As such, this paper contributes an empirical snapshot of the use of interactive computing platforms in ML practices, and argues for a renewed focus in the field of AI ethics on the emergence of digital platform infrastructures in the ML ecosystem.

%% file: sections/2_related_work.tex
\subsection{The sociomaterial context of Machine Learning practices}
\label{subsec:MLcontext}

As ML systems have become objects of sociological interest \cite[e.g.][]{burrell2016how, dourish2016algorithms, kitchin2014big, mackenzie2015production}, the social context in which ML systems are researched, developed, commissioned, and deployed has garnered increased attention in diverse fields from Human-Computer Interaction \cite{holstein2019improving, madaio2022assessing}, to Science and Technology Studies \cite{christin2017algorithms}, to public policy \cite{krafft2020defining}. In this paper, I refer to \textit{sociomaterial} context rather than \textit{social} context to signal a particular focus on the intertwining of socially-constructed material things---specifically, interactive computing platforms---in ML practices. As Paul Leonardi et al. \cite{leonardi2008materiality, leonardi2012materiality, leonardi2012materialitya} have argued, a sociomaterial perspective highlights how the material is socially constructed, and the social is enacted through material forms. A sociomaterial perspective invites us to consider the material things that ML practitioners enrol in their day-to-day work, alongside other aspects of the social context, and the contribution of these things to the stablisation of ML practices. In this context, \textit{material} refers to the ``properties of a technological artifact that do not change, by themselves, across differences in time and context'' \cite[p.7]{leonardi2012materiality}---for interactive computing platforms, and software generally, this includes their user interfaces and layouts, their core capabilities, and their dependencies \cite{redstrom2005technology}. My understanding of \textit{practices} is informed by social practice theory \cite{rouse2007practice, bourdieu2020outline, sloane2022german}, which conceptualises \textit{practices} as routinised ways of understanding and performing social activities \cite{ingram2007products}, and highlights that multiple practices can co-exist within the same cultural setting \cite[p.646]{rouse2007practice}. \textit{Machine Learning practices} are thus the constitutive matter of `doing' ML. Some practices (e.g. Agile meeting processes) may be widely shared across cultures and organisations, and others (e.g. the use of specific software) may vary dramatically from practitioner to practitioner.\looseness=-1

In the field of AI ethics, a sociomaterial perspective has been used to highlight the challenges of translating AI ethics research into ML practices. Michael Veale and Reuben Binns \cite{veale2017fairer}, for instance, studied how statistical measures of fairness can be implemented within the practical constraints of limited access to data on protected characteristics, finding that new institutional arrangements will be necessary to support industry implementation of statistical measures of fairness that depend on access to sensitive data \cite[cf.][]{beutel2019putting, bogen2020awareness}. Veale and Binns argue for future empirical research on the ``messy, contextually-embedded and necessarily sociotechnical'' challenge of building `fairer' ML systems \cite[p.13]{veale2017fairer}. Veale et al. \cite{veale2018fairness} subsequently conducted an empirical study of ML practitioners in public sector organisations and their engagement with ethics issues during ML system development for high-stakes decision making, finding that while practitioners have a high degree of awareness regarding ethical issues, they lack the necessary tools and organisational support to use this awareness in their ML practices. Mona Sloane and Janina Zakrzewski \cite{sloane2022german}, who also situate their work within social practice theory, provide a more expanded overview of AI ethics practices, through an empirical study of the operationalisation of ethics in German AI startups. Sloane and Zakrzewski develop an anatomy of AI ethics practices, which they suggest can be used as a framework to inform improvements to ML system development practices. Relevantly, the anatomy includes `ethics materials', defined as ``concrete objects, processes, roles, tools or infrastructures focused on `AI ethics''' \cite[p.5]{sloane2022german}. Holstein et al. \cite{holstein2019improving} provide further support for the importance of ethics materials, through their empirical study of ML practitioners working in product teams in large technology firms to develop `fairer' ML systems, which found that practitioners lack the tools needed to identify and address ethics issues that arise during ML system development. Finally, Will Orr and Jenny Davis \cite{orr2020attributions} highlight how ML practices include the diffusion of responsibility for ethics during ML system development. Orr and Davis found a ``pattern of ethical dispersion'' amongst practitioners: practitioners perceive themselves to be the inheritors of ethical parameters from more powerful actors (regulators, clients, employers), which their expertise translates into the characteristics of systems they develop, which are then handed over to users and clients, who assume ongoing responsibility \cite[p.7]{orr2020attributions}. These studies, along with other empirical explorations of ML practice \cite[e.g.][]{vakkuri2020current, hopkins2021machine, rakova2021where, ryan2021research, krafft2020defining, kaur2020interpreting} and several workshops focused on the research-to-practice gap \cite{baxter2020bridging, barry2020ethics, szymielewicz2020where}, have prompted calls for better support for practitioners attempting to operationalise AI ethics principles in their ML practices \cite{morley2020what, schaichborg2021four, schiff2021explaining}.

This study complements and inverts these empirical studies of AI ethics in ML practices. Rather than moving from the social to the sociomaterial, this study moves from the material to the sociomaterial. That is, rather than starting with interviews \cite[e.g.][]{orr2020attributions, sloane2022german, holstein2019improving, veale2018fairness, hopkins2021machine, rakova2021where, ryan2021research} or surveys \cite[e.g.][]{vakkuri2020current, krafft2020defining} of practitioners to explore their understanding and operationalisation of AI ethics in ML practices, the study begins with material artefacts that practitioners use and produce in the course of their ML practices, and explores what light these artefacts may shed on the translation of AI ethics to ML practices. A similar approach is followed by Max Langenkamp and Daniel Yue \cite{langenkamp2022how} in their study of open source ML software use, which consists of a review of code repositories on GitHub to establish the breadth of open source use followed by interviews with practitioners to provide further context. That study takes a broad perspective, exploring trends across open source software use. In contrast, this study takes a narrow perspective, exploring how a specific category of software tool is used in ML practices. 

\subsection{Interactive computing platforms and Machine Learning practices}
\label{subsec:interactivecomputingplatforms}

\begin{figure}[ht]
    \centering
    \includegraphics[width=\linewidth]{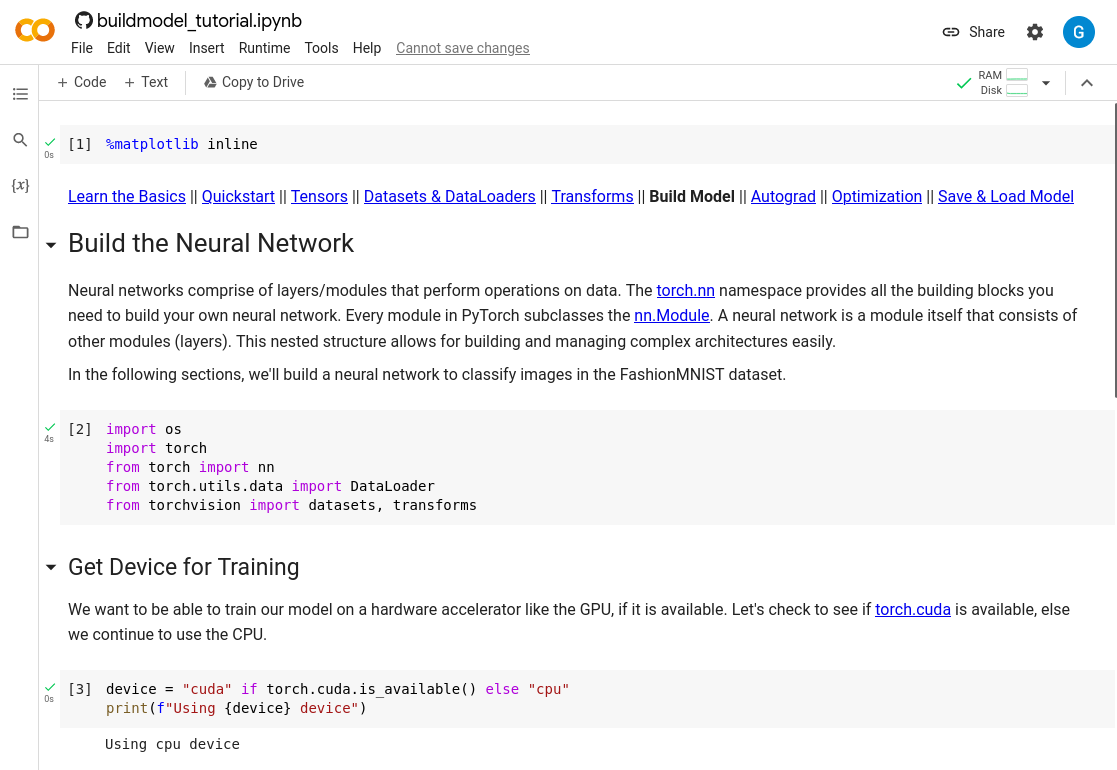}
    \caption{A screenshot of a Google Colab notebook maintained by PyTorch as part of their onboarding documentation. The notebook can be accessed at: \href{https://pytorch.org/tutorials/beginner/basics/buildmodel_tutorial.html}{\url{https://pytorch.org/tutorials/beginner/basics/buildmodel_tutorial.html}}. The numbered grey cells are code cells. Immediately above each code cell is a natural language cell, which contains explanatory text. Immediately below each code cell is the output from running the cell's code.}
    \label{fig:notebook}
\end{figure}

The specific material artefacts this study starts with are \textit{interactive computing platforms} (ICPs), also referred to as `computational notebooks' \cite{rule2018exploration, chattopadhyay2020what}, `literate programming tools' \cite{pimentel2019largescale} or `integrated development environments' \cite{zhang2020how}. Two widely used ICPs are the open-source Jupyter Notebook and Google Colab, Google's extension of Jupyter Notebook, designed to integrate with other Google services.\footnote{Available at \href{https://jupyter.org/}{\url{https://jupyter.org/}} and \href{https://research.google.com/colaboratory/}{\url{https://research.google.com/colaboratory/}}.} Figure \ref{fig:notebook} shows an example Jupyter Notebook. 

Technically, an ICP is an interactive shell for a programming language, such as Python \cite{perez2007ipython}. The shell enables users to write and interact with code fragments---called `cells'---alongside natural language, and to assemble series of cells into a \textit{notebook}, which can be shared with others---much in the same way that a word processor enables a user to assemble an editable document and share that with others. The notebook can be thought of as a computational narrative, which enables one to read and interact with a sequence of code alongside a narrative description of what the code does \cite{kluyver2016jupyter}---hence, the terms `literate programming tools' and `computational notebooks'. However, crucially, an ICP presents itself as only a shell: all but the most rudimentary code fragments depend on access to libraries of existing code, which a user must import into the shell environment. Similarly, particularly in the context of ML, data must be imported into the shell for the code to operate on and a compute resource must be accessed to process operations. Numerous digital infrastructures support importing code into a notebook, including the code repository GitHub,\footnote{See \href{https://github.com/}{\url{https://github.com/}}.} in which many repositories include a notebook to demonstrate common use cases of the code \cite{rule2018exploration}, PyPi,\footnote{See \href{https://pypi.org/}{\url{https://pypi.org/}}.} which indexes and hosts Python-based code packages, and HuggingFace,\footnote{See \href{https://huggingface.co/}{\url{https://huggingface.co/}}.} which indexes and hosts ML training and evaluation datasets and models. In this study, \textit{interactive computing platform} is thus preferred, as the infrastructural implications of `platform' are a critical aspect of what defines these tools: ICPs are highly networked arrangements, one part of a circular web of infrastructures and inter-dependencies (the Internet, cloud computing, programming languages and libraries).

Interactive computing platforms pre-date the widespread adoption of ML techniques in applied settings. Indeed, their motivating design goal was to support reproducible science \cite{kluyver2016jupyter, perez2007ipython, granger2021jupyter} (see, e.g. \cite{beg2021using, randles2017using} for discussions of their effectiveness at meeting this goal). However, as ML techniques have become ubiquitous, and data scientists have become widespread in industry, interactive computing platforms have become widely enrolled in ML practices. Commentators thus describe ICPs as the ``tool of choice'' for data scientists \cite{perkel2018why}, and practitioners vigorously debate the merits and drawbacks of using ICPs in applied settings \cite[e.g.][]{ufford2018beyond, brinkmann2021jupyter, grusdon2018i, mueller2018reasons, howard2020creating}.

Interactive computing platforms have also become objects of study in several fields adjacent to AI ethics. Human-Computer Interaction studies have developed empirical accounts of the way users interact with ICPs, focusing particularly on the role of ICPs in collaborations \cite{zhang2020how, wang2019how} and in data science \cite{rule2018exploration, kery2018story}. Of particular relevance, Adam Rule et al. \cite{rule2018exploration} conducted three studies of the use of ICPs by data scientists, which included a large-scale review of notebooks on GitHub and interviews with data scientists and found that ICPs tend to be used by data scientists during data exploration phases of a project, rather than for constructing and sharing detailed explanations of data analysis. Studies in the field of Software Engineering have also focused on documenting the use of ICPs, focusing particularly on ICPs as a site to study trends in code use \cite{wang2020better} and reuse \cite{koenzen2020code, pimentel2019largescale}, and on the their potential as educational tools \cite{tan2021nascent}. Similarly, in the field of Computational Science, several studies have considered the role ICPs can play in supporting reproducible science \cite{beg2021using, brown2021reproducing, juneau2021jupyterenabled}. This study provides a different perspective on ICP use, by considering ML practices in particular, and interpreting these practices through the lens of sociological studies of infrastructure, which shifts the focus of the study away from the individual user-notebook relationship and towards the broader relationship of ML practitioners to the suite of infrastructures involved in ML practices.

%% file: sections/3_theoretical_lens.tex

Studying the relationship between practices and infrastructures can be vexed. Infrastructures may be functionally invisible to the social groups who make use of them in daily practices \cite{star1999ethnography}, as I consider further in Section \ref{subsec:infra_invisibile}. Further, infrastructures often span multiple practices across different social groups, which, particularly in the context of digital infrastructures, may not be geographically proximate \cite{bowker2009information}. And, practices themselves are not purely infrastructural---as Shove et al. \cite{shove2012dynamics} argue, they bring together infrastructures and other materials, competencies, and ways of knowing.

Sociological studies of infrastructures have orientated themselves around the broad aim of rendering infrastructures, and their sociopolitical commitments, visible \cite{bowker2009information}. Ethnographic methods--historically, fieldwork and participant observation; more recently, multi-site studies--have been used to empirically document infrastructures \cite{silvast2019assemblage}. Star \cite{star1999ethnography}, for instance, advocates studying moments of breakdown in infrastructures, seeing these as instances where infrastructures become visible to social groups. Star \cite{star1999ethnography} also observes that infrastructures are often learned as part of group membership, directing attention to moments of transience in social groups (discussed further in Section \ref{subsec:infra_relations}). However, digital infrastructures present particular challenges: one cannot physically access online communities, and the number of physical sites is at least as large as the user-base of the infrastructure \cite{bowker2009information}.\footnote{Although outside the scope of this paper, an additional emerging challenge is automated personalisation of digital infrastructures, which makes obtaining a general view of the infrastructure challenging \cite{troeger2022sociotechnical}. ICPs do not currently afford personalisation in this way.}

In this study, I build on Star's insights by focusing, as a path towards understanding ICPs and their relationship to ML practices, on moments where ML practitioners are either experiencing ICP breakdowns or limitations in their own ICP capabilities. In particular, and reflecting the challenge of direct observation of digital infrastructure use, the primary data source used are the questions asked by ML practitioners on popular online forums. This is supplemented with analysis of ICP affordances and inter-dependencies. This approach follows in the spirit of other studies of digital infrastructure, such as Plantin et al.'s \cite{plantin2018reintegrating} analysis of the documentation and inter-dependencies of the Figshare platform and Andre Brock's \cite{brock2018critical} analysis of Black Twitter through analysis of Twitter interfaces and user generated content, although the study presented here is narrower in scope.

%% file: sections/4_method.tex
This study consisted of an empirical study of user-generated content on the Stack Exchange forums, supported by a close reading of a small number of exemplars texts \cite{jacobs2019topic}. In particular, a Structured Topic Model (STM) \cite{roberts2013structural, roberts2014structural, roberts2016model, roberts2019stm} of user-generated questions about ML and the use of interactive computing platforms on Stack Exchange forums was estimated.\footnote{See \cite{isoaho2021topic, nikolenko2017topic, lindstedt2019structural, mohr2013introduction} for overviews of topic modelling in the social sciences, and \cite{brookes2019utility, baumer2019speaking} for more critical perspectives.} A similar approach has been used in a number of studies of Stack Exchange forums \cite{ahmad2018survey}, for instance to identify challenges practitioners face in developing ML systems more generally \cite{alshangiti2019why} or themes in questions asked by mobile application developers \cite{rosen2016what} or themes in privacy-related \cite{tahaei2020understanding} or security-related questions \cite{yang2016what}.\footnote{Code to reproduce pre-processing steps and the topic model described below, are available at \href{https://github.com/gberman-aus/aies_23_topic_modelling}{\url{https://github.com/gberman-aus/aies_23_topic_modelling}}.}

\subsection{Corpus development and description}

English-language Stack Exchange community forums, specifically Stack Overflow, Cross Validated, Data Science, Computer Science, and Software Engineering were mined for relevant questions. Stack Exchange claims to be the world's largest programming community.\footnote{See \href{https://stackexchange.com/}{\url{https://stackexchange.com/}} to access Stack Exchange and its forums. Stack Overflow is broadly focused on computer programming. Cross Validated is a more specialised forum, focused on statistics and data analysis. Software Engineering is a similarly specialised forum, focused on software systems development. Finally, Data Science and Computer Science are relatively small forums, focused on data and computer science respectively. However, reflecting the ubiquity of ML techniques in computing, questions related to ML occur in all of these forums, and, as all of these forums are user-moderated, their boundaries and scope are dynamic.} As of October 2022, its most popular forum, Stack Overflow, had over 19 million registered users, who contribute, edit, and moderate questions and answers on the forum.\footnote{This estimate is based on a query of the Stack Exchange Data Dump. See \cite{baltes2019usage, barua2014what} for studies of Stack Overflow usage.} Previous research demonstrates that Stack Overflow is enmeshed in software engineering and data science practices \cite[e.g.][]{treude2019predicting, baltes2019usage, an2017stack, abdalkareem2017code, nasehi2012what, ford2016paradise}, and that ML techniques are a rapidly growing topic of discussion on the forum \cite{alshangiti2019why}. The Stack Exchange forums share data structures\footnote{A detailed description of the database schema used across forums is provided by Stack Exchange on their forum about the Stack Exchange network, appropriately named Meta Stack Exchange, accessible at \href{https://meta.stackexchange.com/questions/2677/database-schema-documentation-for-the-public-data-dump-and-sede}{\url{https://meta.stackexchange.com/questions/2677/database-schema-documentation-for-the-public-data-dump-and-sede}}.} and interface layouts, with annoymised user questions, answers, and comments from all Stack Exchange forum made available for querying and research through the Stack Exchange Data Dump \cite[e.g.][]{yang2016what, rosen2016what, alshangiti2019why}.\footnote{The Stack Exchange Data Dump can be accessed at: \href{https://archive.org/details/stackexchange}{https://archive.org/details/stackexchange}. The database is updated weekly.}

\begin{figure*}
    \centering
    \begin{subfigure}[b]{0.4\textwidth}
        \centering
        \includegraphics[width=\textwidth]{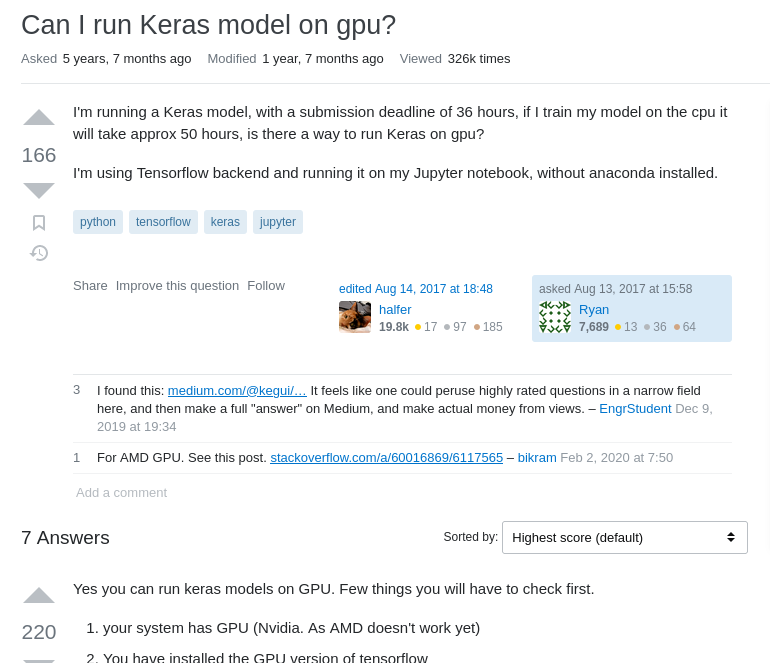}
        \caption{Topics: 13 (28.1\%), 5 (24.5\%), and 23 (13.3\%). The infrastructure and inter-dependencies cluster.}
        \label{fig:keras2_q}
    \end{subfigure} \hfill 
    \begin{subfigure}[b]{0.4\textwidth}
        \centering
        \includegraphics[width=\textwidth]{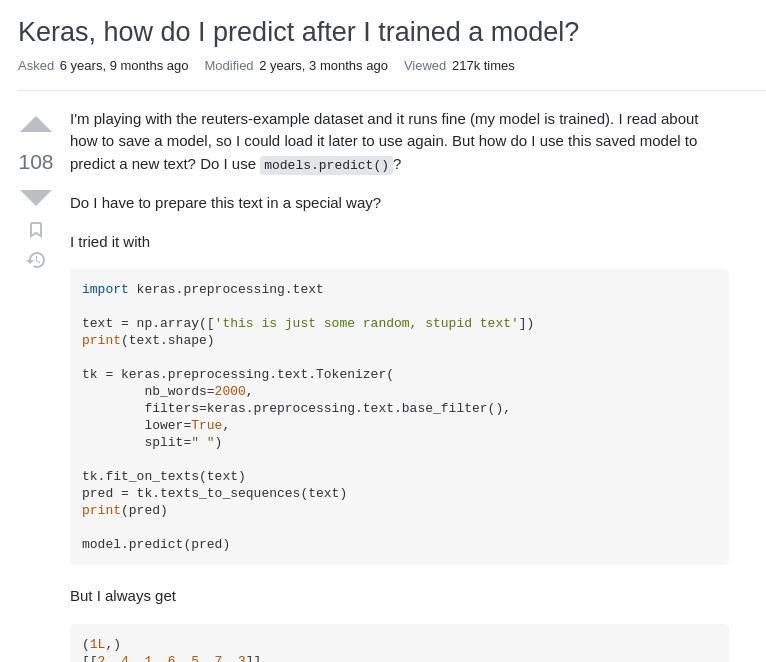}
        \caption{Topics: 19 (15.6\%), 21 (11.2\%), and 16 (10.3\%). The data manipulation cluster.}
        \label{fig:keras_q}
    \end{subfigure} 
    \vskip\baselineskip
    \begin{subfigure}[b]{0.4\textwidth}
        \centering
        \includegraphics[width=\textwidth]{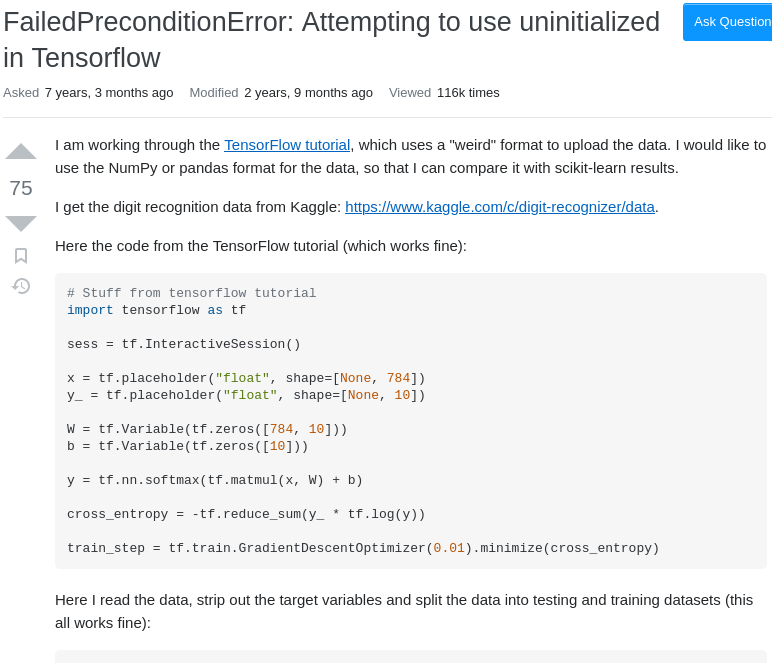}
        \caption{Topics: 25 (48.2\%), 4 (9.6\%), and 8 (7.3\%). The model training cluster.}
        \label{fig:tensor_q}
    \end{subfigure} \hfill
    \begin{subfigure}[b]{0.4\textwidth}
        \centering
        \includegraphics[width=\textwidth]{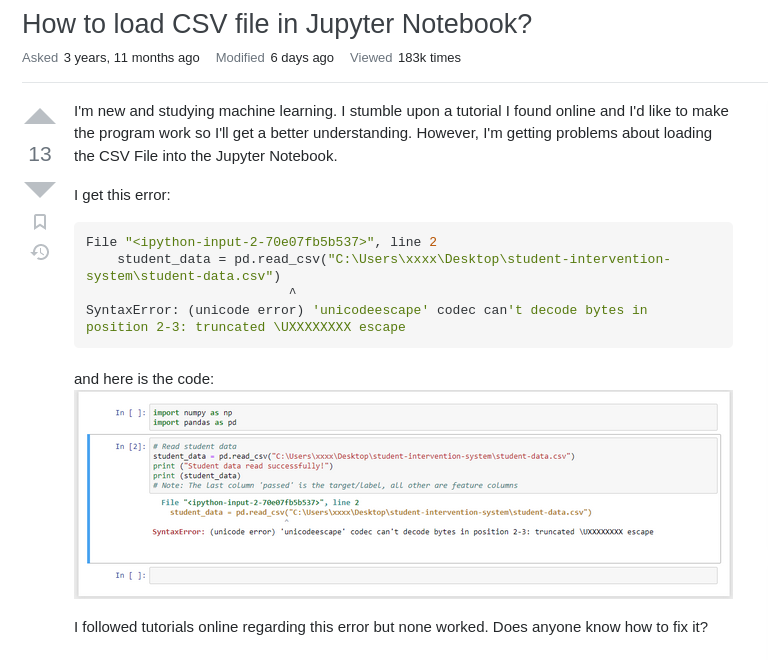}
        \caption{Topics: 13 (44.2\%), 5 (31.4\%), and 12 (6.6\%). The infrastructure and inter-dependencies cluster.}
        \label{fig:notebook_q}
    \end{subfigure}
    \caption{Screenshots of four highly viewed questions on the Stack Overflow forum. The top three topics identified by the topic model and the cluster are reported in the caption of each image.}
    \label{fig:example_questions}
\end{figure*}

Questions related to ML and interactive computing platforms were extracted from the Stack Exchange forums listed above on 23 November, 2022. Four example questions are shown in Figure \ref{fig:example_questions}. To identify relevant questions the topical tags associated with every question were leveraged. Through manual review of the forums, and queries of the Stack Exchange Data Dump, $10$ ICP tags and $32$ ML related tags were identified.\footnote{In studies of more niche topics only one tag has been used \cite{tahaei2020understanding}, however, as in \cite{alshangiti2019why}, manual review demonstrated that there are no over-arching ML or ICP tags.} These tags are listed in Appendix \ref{subsec:dumptags}. Having identified relevant ML and ICP tags, two datasets were extracted from the Stack Exchange Data Dump: all questions on the selected forums with at least one ML related tag (a large dataset consisting of $485,053$ questions), and all questions on these forum with at least one ICP related tag (a smaller dataset of $75,639$ questions). The ML tagged questions were filtered by the presence of an ICP term (leaving $36,940$ questions), and ICP tagged questions were filtered by ML terms (leaving $9,634$ questions). This procedure resulted in two datasets with some substantial overlap. After de-duplication, a final dataset of $21,555$ ML and ICP related questions was left; this dataset became the corpus used to estimate a STM topic model..\footnote{A significant advantage of the $stm$ R package relied upon is that it enables manual setting of the random seeds used during the model training process---ensuring a higher degree of reproducability is possible.}

\subsection{Estimation of the topic model}
\label{subsec:estimation}

STM is a probabilistic, mixed-membership topic model, which extends the widely-used Latent Dirichlet Allocation model by enabling the inclusion of metadata---here, the tags associated with questions and question creation date---in the model training process (see \cite{lindstedt2019structural, roberts2014structural} for introductions to STM). To prepare the corpus for topic modelling, pre-processing was undertaken using the $stm$ R package \cite{roberts2019stm} (see \cite{grimmer2022text, wesslen2018computer} for discussion of pre-processing procedures). Title and body fields for questions were concatenated into a single column. Questions on Stack Exchange forums are formatted using markdown, and often include large snippets of computer code. All code snippets and markdown were removed from questions. Code snippets were retained for subsequent analysis. Html symbols (e.g. `\&quot;'), special characters (e.g. `\&\#39;'), punctuation, and superfluous white spaces were removed from questions. Questions were converted to lowercase. Frequently occurring words with little topic predictive value ('stopwords') were removed from questions. Words in the questions were stemmed (i.e. converted to their root form). The creation date of questions was converted into a numerical format.

STM requires the researcher to set the number of latent topics ($k$) to identify in a corpus. As such, selecting the optimal value for $k$ is an important decision, and requires testing a wide range of values \cite{grimmer2013text}. Additional hyper-parameters can also be optimised, and different pre-processing regimes can also be tested against each other \cite{grimmer2022text, maier2018applying}. Given the preliminary nature of the study, $k$ values from $10$ to $60$, at intervals of $5$ were experimented with. The $stm$ package's built in multi-model testing feature was used: for each value of $k$, up to $50$ model runs, with a maximum of $100$ iterations each, were tested to ensure model stability. 

To select an optimal value of $k$ two evaluation metrics were used: \textit{exclusivity} and \textit{semantic coherence} \cite{roberts2014structural}. \textit{Exclusivity} is a measure of the difference in key words associated with each topic, whilst \textit{semantic coherence} is a measure of how internally consistent each topic's key words are \cite{wesslen2018computer}. These measures tend to pull in opposite directions: exclusivity is likely to be optimised by increasing the number of topics, whilst semantic coherence can be optimised by decreasing the number of topics \cite{roberts2014structural}. An optimal number of topics for social science research can be found by plotting exclusivity against semantic coherence for a range of $k$ values, and then choosing a value at which neither measure dominates \cite{roberts2014structural}. However, there is no `right' value for $k$ \cite{quinn2010how}; the aim is to find a value of $k$ that enables meaningful interpretation \cite{grimmer2013text, roberts2014structural, syed2017full}. In this instance, as can be seen in Figure \ref{fig:excl_v_coherence} in the Appendix, models with $k$ values of around $25$ represented an optimal trade off between exclusivity and semantic coherence. After inspection of keywords associated with each topic and representative questions, the model with a $k$ value of $30$ was selected.\looseness=-1

\subsection{Interpretation of the topic model}

To interpret the results of the topic model Yotam Ophir and Dror Walter's \cite{ophir2020collaborative, walter2019news} three step process was followed. First, I qualitatively interpreted the topics identified through review of the most probable words associated with each topic (Figure \ref{fig:k30_summary}). Second, I analysed the relationships between topics by calculating their correlation, with a positive correlation indicating a high likelihood of two topics being found together in the one Stack Exchange question \cite{roberts2016model, roberts2019stm}. Third, I used a community detection algorithm to identify clusters of topics and broader themes across the corpus (Figure \ref{fig:topic_network}). In particular, I used the Newman-Girvan method for community structure detection \cite{newman2004finding}, with the result being three clusters of topics. After review of the probable terms associated with topics within each cluster and representative questions, I labeled these clusters: \textit{infrastructure and inter-dependencies}, \textit{data manipulation}, and \textit{model training}. As an additional final step, I made use of STM's ability to calculate the impact of covariates on topic prevalence to analyse the expected proportion of individual topics (Figure \ref{fig:13_15_time}) and clusters of topics over time (Figure \ref{fig:network_time}). 

Throughout the above steps I moved between analysis of the topic model itself and deeper review of full Stack Exchange question and answer threads that are representative of particular topics or clusters of topics. Here, I adapted the approach of Paul DiMagggio, Manish Nag, and David Blei \cite{dimaggio2013exploiting}, who, after training a topic model, identify topics of interest and then undertake analysis of the most representative texts for those topics. In particular, I identified the $10$ Stack Exchange questions with the highest probability for each topic, and the $10$ questions with the combined highest average probability across topics within each cluster. For these highly representative questions, within each topic cluster I further sorted the questions by their view count on the Stack Exchange forums (Figures \ref{fig:cluster_quotes_1} - \ref{fig:cluster_quotes_3} in the Appendix), enabling me to identify questions that were both highly representative of a given topic cluster and highly viewed on the Stack Exchange forum.

%% file: sections/5_findings.tex
\begin{figure*}
    \centering
    \begin{subfigure}[b]{0.4\textwidth}
        \centering
        \includegraphics[width=\linewidth]{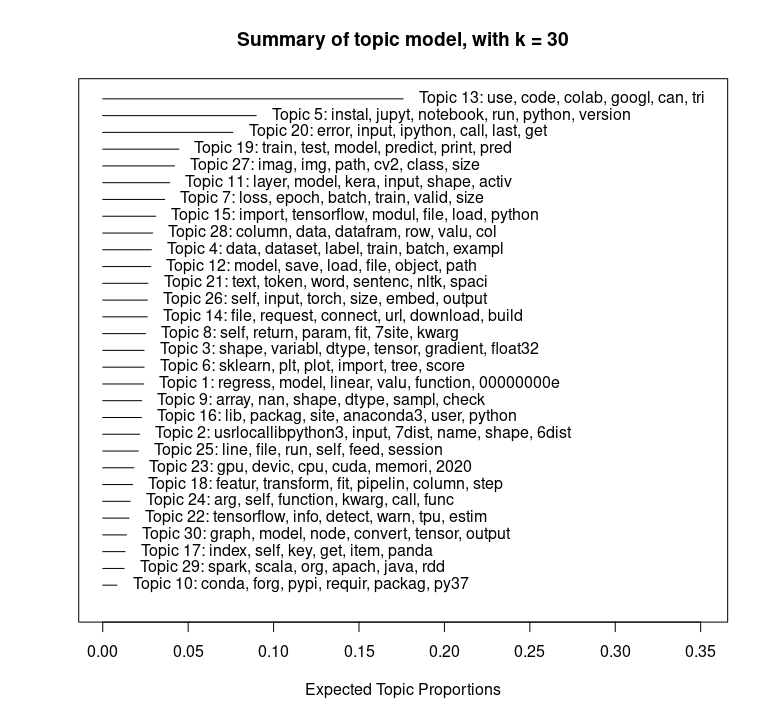}
        \caption{Expected distribution of all topics across the corpus, with the most probable word associated with each topic.}
        \label{fig:k30_summary}
    \end{subfigure} \hfill
    \begin{subfigure}[b]{0.4\textwidth}
        \centering
        \includegraphics[width=\linewidth]{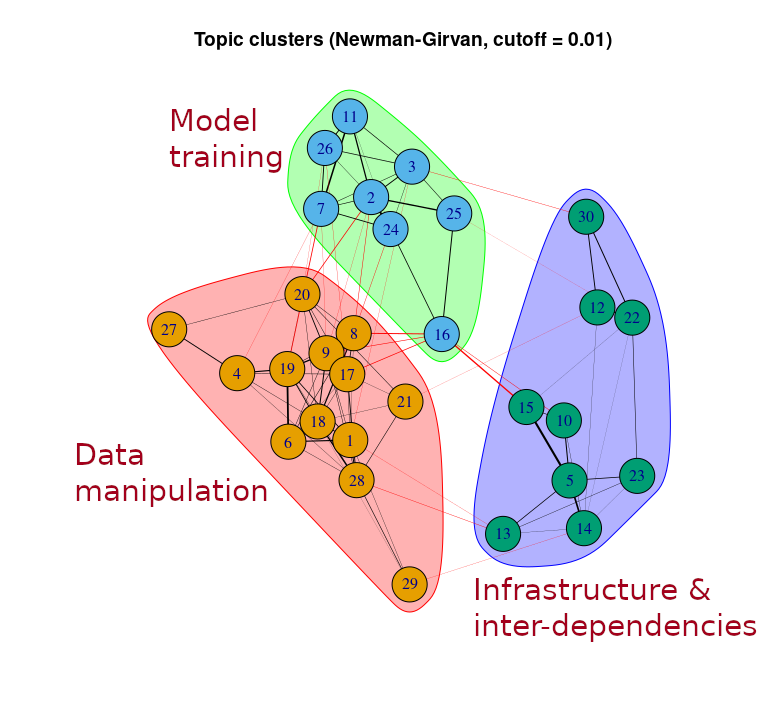}
        \caption{Topic correlation network, using the Newman-Girvan method, with a minimum correlation threshold of 0.01.}
        \label{fig:topic_network}
    \end{subfigure}
    \vskip\baselineskip
    \begin{subfigure}[b]{0.4\textwidth}
        \centering
        \includegraphics[width=\linewidth]{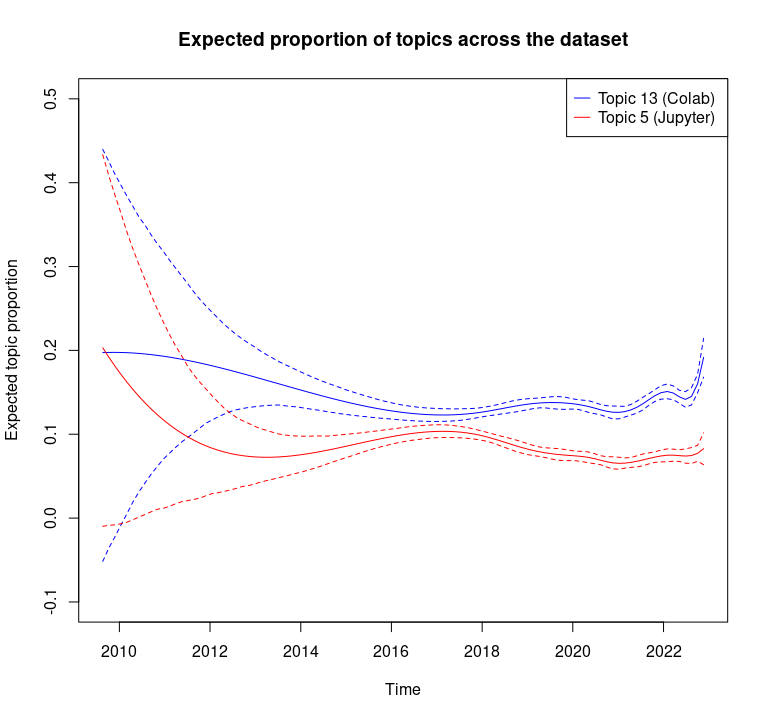}
        \caption{Comparison of the expected topic proportions in the corpus over time for topics 13 (Colab related questions) and 5 (Jupyter related questions). Dashed lines represent a confidence interval of 0.95.}
        \label{fig:13_15_time}
    \end{subfigure} \hfill
    \begin{subfigure}[b]{0.4\textwidth}
        \centering
        \includegraphics[width=\textwidth]{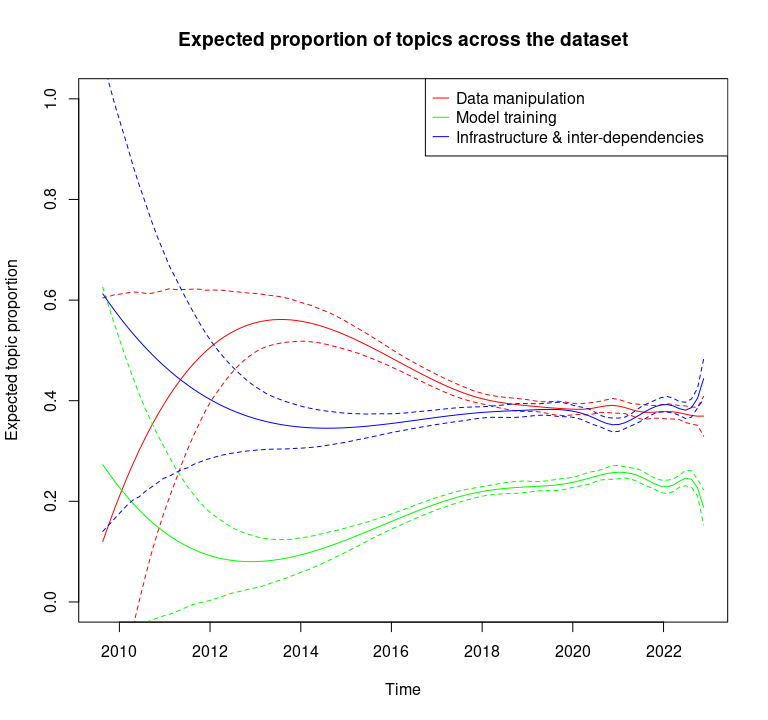}
        \caption{The expected topic proportions over time, with the three communities identified in Figure \ref{fig:topic_network} treated as groups of topics. Dashed lines represent a confidence interval of 0.95.}
        \label{fig:network_time}
    \end{subfigure}
    \caption{Visualisations of the estimated topic model.}
    \label{fig:model_visuals}
\end{figure*}

The topic model of Stack Exchange questions discussing interactive computing platforms and ML demonstrates that interactive computing platforms are implicated in a wide range of ML practices. ML practices are often conceptualised within a life cycle framework, with stages of ML development moving from problem formulation, to data curation and processing, to model training and evaluation, to model deployment and ongoing monitoring \cite[e.g.][]{ashmore2022assuring, lee2021risk, polyzotis2018data, polyzotis2017data, morley2020what}. Figure \ref{fig:k30_summary} shows the most probable terms associated with each topic, and the expected proportion of each topic across the corpus. Unsurprisingly, given the corpus focus on ICPs, the two topics most widely represented in the dataset---$13$ and $5$---are associated with Google Colab and Jupyter Notebook respectively. The most probable terms for most other topics are associated with many of the ML development stages, particularly data curation and processing (e.g. see key terms for topics $28$, $4$, and $21$), and model training and evaluation (e.g. see key terms for topics $19$, $11$, $7$, and $12$). The deeper review of identified topics and representative questions highlights two inter-related themes, which address the study's research question regrading use of ICPs in ML practices: the use of ICPs as \textit{learning laboratories}; and, their role as \textit{coordination hubs} across ML infrastructures.\looseness=-1

\subsection{Learning laboratories for Machine Learning}
\label{subsec:code_reuse}

Interactive computing platforms serve as ML practice \textit{learning laboratories}: they enable users to experiment with each other's code and publicly-available datasets, learn how code functions through line-by-line interactions, and redeploy code in their own use cases. ICPs are thus part of the sociomaterial context for what Louise Amoore has described as the ``\textit{partial, iterative and experimental}'' nature of ML practices \cite{amoore2019doubt}, which is also reflected in Langenkamp and Yue's broader study of open source tools \cite{langenkamp2022how}.\footnote{For an extended description of the relationship between learning practices and digital infrastructures see \cite{guribye2015artifacts}.}

Figure \ref{fig:notebook} shows an example of an ICP used as a learning laboratory, drawn from a tutorial for PyTorch, an ML-focused high-level programming language. Figure \ref{fig:keras_q} shows an example of a Stack Overflow question, titled `\textit{Keras, how do I predict after I trained a model?}'', which also reflects the use of an ICP as a learning laboratory. This is one of the four most viewed questions from the data manipulation cluster of topics. The author of this question appears to be engaged in a learning practice: they describe themselves as ``\textit{playing with}'' the dataset, and write that they have ``\textit{read about}'' saving a trained model, but are now struggling to use the saved model in a prediction task. Not shown in Figure \ref{fig:keras_q} are the community answers the author received.\footnote{The full question, including community provided answers can be seen as: \href{https://stackoverflow.com/questions/37891954/keras-how-do-i-predict-after-i-trained-a-model}{\url{https://stackoverflow.com/questions/37891954/keras-how-do-i-predict-after-i-trained-a-model}}.} Each answer also includes a code snippet, demonstrating a solution. Similarly, the question ``\textit{FailedPreconditionError: Attempting to use uninitialized in Tensorflow}'' (Figure \ref{fig:tensor_q}), one of the most viewed questions in the model training cluster, includes a code snippet that is ``\textit{from the TensorFlow tutorial}'', which the author is attempting to use with ``\textit{digit recognition data from Kaggle}''. In both these questions users' learning is through an ICP, and is focused on understanding how to achieve a specific task using the Application Programming Interface (API) of a particular high-level programming library. 

When ML practitioners use interactive computing platforms as learning laboratories they engage in practices of code and data reuse. The author of the Stack Overflow question discussed above notes they are ``\textit{playing with the reuters-example dataset}'', which is a publicly-available dataset used in topic modelling and text classification tasks \cite{lewis1997reuters}, and provides a code snippet to illustrate the point at which they require assistance. Within ML practices reuse of publicly available datasets, such as the Reuters dataset for text classification or the ImageNet dataset for computer vision is well documented \cite{denton2021genealogy}. Patterns of dataset reuse can be found across the corpus: the Reuters dataset is referend in $11$ questions; ImageNet dataset is mentioned in $436$ questions; and, the MNIST handwritten digits dataset is mentioned in $834$. Indirect evidence of code and data reuse in ML practices can also be found by reviewing the code snippets included in questions in the corpus. As discussed in Section \ref{sec:method}, during pre-processing code snippets were isolated from the text of questions on which the topic model was trained. Of all questions, 89.7\% include a code snippet. Because the corpus consists of questions about using ICPs, many of these code snippets represent the point at which a user of an ICP has become stuck while trying to attempt to an ML related task. This is illustrated by the question titled ``\textit{How to load CSV file in Jupyter?}'', shown in Figure \ref{fig:notebook_q}. Here, the author of the question has included in the body of their question a screenshot of their Jupyter Notebook. As can be seen, the first cell in this notebook begins with the $import$ function, which is how specific programming libraries or sub-libraries are imported into the ICP. In this case, the author has imported $numpy$, a mathematical functions library, and $pandas$, a data analysis library. More broadly, the code snippets included in questions shed light on the substance of code that is entered into ICPs during ML related tasks. By calculating the frequency of the terms that immediately follow the $import$ function, widely used programming libraries can be identified (see Figure \ref{fig:libraries} in the Appendix). Among the $15$ most frequently mentioned programming libraries in code snippets are: `Sequential', `Dense', and `Model' (specific components from Keras, a high-level library for deep learning); 'cv2' (a computer vision high-level library); and, `PyTorch' (an alternative to TensorFlow).\looseness=-1

The code snippet in the question titled ``\textit{FailedPreconditionError: Attempting to use uninitialized in Tensorflow}'', shown in Figure \ref{fig:tensor_q}, illustrates the significance of $import$ functions for extending the abilities of ICPs both as learning laboratories and more generally. The code snippet includes the line: \newline
\begin{center}
    $train\_step = tf.train.GradientDescentOptimizer$\newline
\end{center}
Across the corpus, $120$ questions reference TensorFlow's \textit{GradientDescentOptimizer}. Gradient Descent is a type of optimisation algorithm used during training of a neural network \cite{ruder2016overview}. This line of code enables the user to access the TensorFlow library's operationalisation of gradient descent algorithms through its API---alleviating the need for the user to code their own gradient descent algorithm. While TensorFlow is only one of a number of similar software libraries available, the volume of posts ($38.6\%$ of all questions) in the corpus in which TensorFlow is mentioned, and the two most probable terms in topic $15$ (`import' and `tensorflow'), provides some indication of its widespread use in ICPs and ML practices.\looseness=-1

\subsection{Coordination hubs for ML infrastructures}
\label{subsec:infrastructures}

Assembling an ML workflow is a complex task, requiring coordination of multiple infrastructures. Interactive computing platforms serve as \textit{coordination hubs}, through which networks of infrastructures are assembled to support ML practices. Reflecting this, as shown in Figure \ref{fig:network_time}, the cluster of topics associated with infrastructure and inter-dependencies accounts for a significantly greater proportion of questions in the corpus than the cluster of topics associated with model training. The most viewed questions within the infrastructure and inter-dependencies cluster reveal the infrastructural coordination that is at the heart of many ML practices.

One of the most viewed questions within the infrastructure and inter-dependencies cluster is titled ``\textit{Can I run Keras model on gpu?}'' (Figure \ref{fig:keras2_q}).\footnote{See \href{https://stackoverflow.com/questions/45662253/can-i-run-keras-model-on-gpu}{\url{https://stackoverflow.com/questions/45662253/can-i-run-keras-model-on-gpu}} for the full question and its answer thread.} Keras is a high-level API designed to support deep learning techniques.\footnote{See \href{https://keras.io/}{\url{https://keras.io/}} for an introduction to Keras.} Keras is integrated into TensorFlow, and enables users to build a wide range of neural networks---Keras makes it easier and more efficient to complete deep learning tasks within TensorFlow. A GPU---Graphics Processing Unit---is a specialised microprocessor, which in many computing systems works alongside the more general-purpose Central Processing Unit (CPU) microprocessor. Whilst the GPU-CPU arrangement predates the emergence of Deep Learning, it turns out that GPU microprocessors are better suited to performing many of the computations required to train a neural network than CPUs. The author of this question is attempting to assemble a system that consists of a ``\textit{Tensorflow backend}'' and a ``\textit{Keras model}'', interacted with through a ``\textit{Jupyter notebook}'', and run on their computer's GPU. The highest scoring answer recommends installing CUDA, which is an additional parallel programming platform designed to enable GPUs to be used for non-graphics processing tasks, such as model training. This answer provides hyperlinks to additional resources for installing CUDA and checking that TensorFlow is running properly on a GPU. Above this answer are two further user comments also linking to additional resources. As such, the author of this question is assembling a system that involves at least five interdependent layers: GPU, CUDA, TensorFlow, Keras, Juypter Notebook. The author is fortunate, however, as their aim is to train their model within ``\textit{36 hours}'', which suggests that either they have access to a powerful GPU, or they are training a model with a relatively small dataset (for instance, as part of a learning exercise). In industrial or research settings, training a neural network requires access to much greater compute resources, which requires users to access a cloud resource, such as Amazon Web Services, and adds at least one additional layer of complexity to the system.

The key words associated with topics within the infrastructure and inter-dependencies cluster (shown in Figure \ref{fig:k30_summary}) provide an additional perspective on the infrastructural coordination required to support ML tasks. In descending order of representation in the corpus, these topics are: $13$, $5$, $15$, $12$, $14$, $23$, $22$, $30$, and $10$. As already observed, topics $13$ and $5$ relate to Google Colab and Jupter Notebook, two ICPs. Meanwhile, topic $15$ includes `import' and `tensorflow' as the two most probable terms. Topic $23$ includes `gpu', `cpu', and `cuda' as probable terms. Topic $22$ includes `tensorflow' and `tpu', which is a reference to Tensor Processing Units, which are a new generation of GPUs specifically designed to support TensorFlow. The presence of these topics, and their close correlations, as shown in Figure \ref{fig:topic_network}, indicate that coordination between infrastructures is widely discussed on Stack Exchange. Finally, Figure \ref{fig:network_time} shows the expected proportion of topic $13$ (Google Colab) compared to topic $5$ (Jupyter Notebook) over time. The topic model estimates that since 2017 questions related to Google Colab have increased compared to questions related to Jupyter Notebook. Significantly, a key point of difference between these two platforms is that Google Colab has been designed to integrate directly into Google's cloud compute infrastructure, and is used as the platform of choice in TensorFlow and Keras tutorials.

%% file: sections/6_discussion.tex
In this study, the intertwining of interactive computing platforms in ML practices was explored. The findings indicate that ICPs are learning laboratories---tools by which users experiment with and learn ML practices through line-by-line interaction with others' code and publicly available datasets, facilitated by the APIs provided by high-level programming languages. The findings show also that ICPs are coordination hubs---sites at which multiple different infrastructures are brought together to support ML practices, such as model training or data processing. Given the role ICPs play as coordination hubs for ML practices, they can be conceptualised as an emerging form of `digital infrastructure'---an essential and widely participated in sociotechnical system \cite{plantin2019digital, bowker2009information}. Conceptualising ICPs in this way enables existing theorising about infrastructures to inform consideration of the sociopolitical significance of ICP use in ML practices, and helps connect ICP use to concerns raised in AI ethics discourse. To illustrate this, in the following subsections I consider how Brian Larkin's review of anthropological practices for studying infrastructure \cite{larkin2013politics} and Susan Leigh Star's description of the properties of infrastructure \cite{star1999ethnography} can apply to ICPs. In each subsection I conclude with a brief reflection on implications for AI ethics discourse.\looseness=-1

\subsection{An emerging infrastructural relationship}
\label{subsec:infra_relations}

As material objects, Larkin describes infrastructures as ``\textit{built networks that facilitate the flow of goods, people, or ideas}'' \cite[p.328]{larkin2013politics}. At the same time, infrastructures are systems that support the functioning of other objects, and it is these objects that users of an infrastructure experience; we experience hot water, not plumbing \cite[p.329]{larkin2013politics}. Star describes this characteristic of infrastructure as `transparency': for users of an infrastructure, the tasks associated with it seem easy and straightforward---transparent \cite{star1999ethnography}. Star, however, understands infrastructures as relational. Transparency is not an inherent characteristic of a sociotechnical system, but rather a characteristic of an infrastructural relationship between a sociotechnical system and its users. 

The topic model of Stack Exchange questions is a snapshot of an emerging infrastructural relationship: as ML practitioners use ICPs as coordination hubs, facilitating flows of data and code across networks of disparate resources (compute capacity, programming languages, datasets, etc.), they are forming an infrastructural relationship with the platform. The platform itself recedes into the background and the objects that the platform enables to function---predictive models---come into the foreground. This is why ICPs excel as learning laboratories for ML. The affordances of the ICP, however, continue to have efficacy even as the platform itself becomes transparent: the affordances enable and constrain users, and in doing so help configure practices associated with the ML techniques that the platform enables \cite{davis2016theorizing, davis2020artifacts}. 

Star's understanding of infrastructures as relational also highlights the relationship between infrastructures and groups: infrastructures are ``\textit{learned as part of membership}'' \cite[p.381]{star1999ethnography}. This conceptualisation of the relationship between infrastructures and group membership appears to align closely to the burgeoning infrastructural relationship between ML practitioners and ICPs. As the topic model interpretation illustrates, ML practitioners learn to use an ICPs as part of the process of becoming `ML practitioners'. Star highlights that shared use of common infrastructures among practitioners helps reinforce their identity as a distinct group \cite{star1999ethnography}. Non-members, meanwhile, encounter infrastructures as things they need to learn to use in order to integrate into a group. Note, for example, the author's phrasing in the Stack Overflow question shown in Figure \ref{fig:notebook_q}: ``\textit{I'm new and studying machine learning... I'm getting problems about loading the CSV File into the Jupyter Notebook}''. `ML practitioner' is an ill-defined term frequently used in AI ethics discourse as a catchall for describing the data scientists, software engineers, and product managers who work on the research and development of ML systems. From an infrastructural perspective, however, the term can also be thought of symbolising a new set of infrastructural relations: where previously data scientists, software engineers, etc., each worked within their own suites of tools, increasingly they use shared infrastructure, such as ICPs, enabling the collapsing of distinctions between these professional roles that is indicated in the term `ML practitioner'. 

\subsubsection{Implications for AI ethics}

A stream of AI ethics research has focused on the development of software and management tools to support ML practitioners (see \cite{morley2020what} for an overview). For this stream, ICPs may be a constraint, in so far as tool adoption is often held to be dependent on integration with existing ML infrastructure and practices \cite[e.g.][]{hardt2021amazon, gebru2021datasheets}. Alternatively, the affordances of ICPs may offer new opportunities for future tool development. The grammar of ICP interactions may be applied to the design of tools intended to prompt practitioner reflection. The open source Fairlearn library, for example, provides example ICP notebooks\footnote{See \href{https://fairlearn.org/v0.8/auto_examples/index.html}{\url{https://fairlearn.org/v0.8/auto_examples/index.html}}.} to demonstrate library uses.

More broadly, however, as ICPs contribute to the configuring of ML practices, they shape the space in which AI ethics are situated. Here, Britt Paris's \cite{paris2021time} reflections on the relationship between Internet infrastructure and constructions of time are instructive. ICPs, like the Internet at large, imagine particular temporal relations. ICPs, in particular, are premised on speed: the staccato call-and-response of user inputs and computer outputs helps configures a working environment in which the value of ML practices resides in their speed and efficiency. In this sense, conceptualising ICPs as ML infrastructure presents as a challenge to calls from AI ethics researchers for greater reflexivity in ML \cite[e.g.][]{fish2021reflexive, weinberg2022rethinking}.

\subsection{Visible and invisible infrastructures}
\label{subsec:infra_invisibile}

As material objects, infrastructures are designed, and reflect, at least in part, the intentions of the designer. Yet, at the same time, infrastructures are ``\textit{built on an installed base}'', often following paths of development laid down by preceding infrastructures \cite[p.382]{star1999ethnography}. And, infrastructures are often caught in circular webs of relations with other infrastructures: computers rely on the electricity grid to function, and the functioning of the modern electricity grid is reliant on computers \cite{larkin2013politics}. Infrastructures therefore cannot be understood in isolation, in the same way that they cannot be designed in isolation. The role ICPs play as coordination hubs reflect this: they are built on top of the networked and decentralised infrastructures of the Internet, programming languages, and computing. In doing so, ICPs augment and extend these pre-existing infrastructures, both following path dependencies established by these infrastructures and charting new paths for future infrastructures \cite[cf.][]{winner1980artifacts}.\looseness=-1  

Larkin highlights that infrastructures also serve a `poetic' function \cite{larkin2013politics}. Larkin draws on linguist Roman Jakobson's concept of poetics \cite{jakobson1960linguistics}, which holds that in some speech acts the palpable qualities of speech (roughly, sound patterns) have primary importance over representational qualities (i.e. meaning). Infrastructures, argues Larkin, can have a poetic function, not reflected in the declared intentions of designers, nor in their technical capabilities \cite{larkin2013politics}. Researchers of infrastructure, then, must take seriously the aesthetic aspects of infrastructure, and consider how infrastructures not only reflect the declared intention of those who build them, but also their (undeclared) interests. Larkin's description of the poetics of infrastructure mirrors Jenna Burrell's critique of blithe descriptions of algorithms as 'opaque', which ignore the ways the appearance of opaqueness in an algorithmic system can reflect the politics of the institutions who operate them \cite{burrell2016how}. In this context, a significant line of future inquiry pertains to the different politics and interests reflected in the two ICPs identified as widely used by the topic model: Jupyter Notebook and Google Colab. 

For Larkin, the aesthetic aspects of infrastructure include the way infrastructures may at times appear transparent or invisible \cite{larkin2013politics}. Here, Larkin takes issue with Star's description of infrastructures as `invisible'. Star describes this characteristic of infrastructure as "\textit{becoming visible upon breakdown}" \cite[p.382]{star1999ethnography}. By standardising interactions between material objects, users, and other infrastructures, infrastructures become transparent to users, and, when this transparency becomes routine, the infrastructure itself appears invisible. Questions asked on Stack Exchange can thus be interpreted as instances of ML infrastructure becoming visible. To Larkin, however, the claim that infrastructures are invisible can only ever be partially valid: what the affordances of infrastructures make visible and invisible is both an outcome a system's technical capabilities and its poetic functions. Larkin and Star's debate on invisibility thus helps shed light on the mechanism by which ICPs become implicated in the characteristics of ML systems that are developed through their use. As infrastructural systems, ICPs standardise a particular form of presenting and interacting with code---the `notebook' layout of descriptive and computation cells described in Section \ref{subsec:interactivecomputingplatforms}---and this standardisation renders some aspects of ML system development more visible to ML practitioners than others.

Shifts in the aspects of ML system development that are transparent to ML practitioners can have significant impacts on practitioners' understanding of ML technologies. As discussed in Section \ref{subsec:code_reuse}, ICPs support iterative experimentation with the APIs of high-level programming languages, which often occurs through probing and re-purposing of code written by others. Iterative experimentation with the API of a high-level programming language, however, is unlikely to reveal the full range of decisions that the creators of an API have made in operationalising a particular ML algorithm or technique. The point of Keras' \textit{Tokenizer} function (shown in the code snippet in the Stack Overflow question in Figure \ref{fig:keras_q}) is that it enables users to convert the text in a corpus into a series of integers (`embeddings'), so that computations (e.g. topic modelling) can be run on the corpus. The function enables users to choose whether or not to convert text to lowercase, but because the function has a default setting, this choice is not necessary---by default any call of the \textit{Tokenizer} function will convert text to lowercase before conversion to numerical form.\footnote{See \href{https://www.tensorflow.org/api_docs/python/tf/keras/preprocessing/text/Tokenizer}{\url{https://www.tensorflow.org/api_docs/python/tf/keras/preprocessing/text/Tokenizer}} for the \textit{Tokenizer} documentation.} This may seem inconsequential, but it can have a significant downstream impact: converting a corpus to lowercase means that the verb `stack' and proper noun `Stack' will be embedded as semantically identical. 

As APIs of high-level programming languages become more sophisticated, particularly as they start to incorporate pre-trained models for common ML tasks (e.g. image classification, object detection and labelling, sentiment detection), the choices obfuscated by the API become more consequential. The TensorFlow Object Detection notebook\footnote{Accessible at \href{https://www.tensorflow.org/hub/tutorials/tf2_object_detection}{\url{https://www.tensorflow.org/hub/tutorials/tf2_object_detection}}.} uses a CenterNet pre-trained model which was trained on the Common Objects in Context dataset \cite{lin2015microsoft}. This dataset includes labels for $91$ categories of objects, including `plate', `cup', `fork', `knife', `spoon', and `bowl' (but not, for instance, `chopstick'), and it is these objects that the CenterNet model can detect in images. This sequence of choices, and the constraints each choice imports into the ML system, are not surfaced by experimentation with the API in an ICP; the infrastructural relationship between ML practitioners and ICPs renders transparent code reuse, but leaves detailed code knowledge opaque.\looseness=-1

\subsubsection{Implications for AI ethics}

At stake in AI ethics discourse are questions of legitimacy. Arising from the recognition that code operationalises and reifies particular interpretations of essentially contested social constructs \cite{jacobs2021measurement, moss2022objective, bowker2000sorting} is the challenge of locating where and how coding decisions are currently made, and where they ought to be made. What, if any, categories of gender ought to be included as labels in an image dataset \cite{keyes2018misgendering}? High-level APIs, interacted with through ICPs, obscure these decisions, and in doing so further entrench them in ML practices: what is unknown to ML practitioners is unquestioned.  In this sense, the infrastructural relationship between practitioners and ICPs is an example of social arrangements helping configure ML  practices as 'black boxes' \cite{burrell2016how}, and is thus a new challenge to the efforts of AI ethics researchers to embed accountability for decision making in ML development \cite{cooper2022accountability}.\looseness=-1


\subsection{Development of infrastructures over time}

The role of coordination hub lends ICPs and the web of other infrastructures they are related to (compute resources, code repositories, etc.) a semblance of hierarchical coherence. But, while infrastructures may be presented as coherent, hierarchical structures, they are rarely built or managed in this way. Indeed, Jupyter Notebook began life as an open-source project focused on scientific computing within the Python programming language, before being adopted and adapted by ML practitioners and industry \cite{granger2021jupyter}. In this sense, the emergence of ICPs as infrastructure reflects a familiar process of adaptation and translation \cite[cf.][]{hughes1987evolution}. Relevantly, Star highlights that infrastructures are fixed in modular increments \cite[p.382]{star1999ethnography}, with ``\textit{conventions of practice}'' co-evolving alongside the development of the infrastructure itself \cite{star1999ethnography}. Here, Elizabeth Shove's work on the co-evolution of infrastructures and practices offers a potential framework by which to explore how ICPs and ML practices co-evolve \cite{shove2003comfort, shove2016matters}.

Watson and Shove argue that infrastructural relations and practices co-evolve through processes of aggregation and integration \cite{watson2022how}. Aggregation refers to ``\textit{the ways in which seemingly localised experiences and practices combine and, in combining, acquire a life of their own}'' \cite[p.2]{watson2022how}. Individual ML practitioners, for example, each develop their own approach to coordinating the different layers of infrastructure needed to support ML tasks. However, as individuals share their approaches, and these coalesce into conventions, the conventions themselves shape future infrastructure development. The convention of using GPUs for model training, for example, creates the demand needed to justify the development of more specialised TPUs. Integration refers to ways that policies, processes, and artefacts at the level of the overarching infrastructure are ``\textit{brought together in the performance of practices enacted across multiple sites} \cite[p.2]{watson2022how}. Google, for example, sets various policies regarding the availability of Google's GPU resources to users of Google Colab. These policy decisions (e.g. the decision to offer limited free access to GPUs) in turn are integrated into individual users' ML practices---top-down policy decisions help inform the future development of conventions of practice, but do not determine them. For the field of AI ethics, the framework of aggregation and integration offers a path towards understanding how norms in ML practices, such as the use of particular operationalisations of fairness metrics, co-evolve as a product of both the integration of particular fairness approaches into high-level programming languages and the aggregation of local approaches to `managing' ethics issues into shared practices.

\subsubsection{Implications for AI ethics}

Conceptualising interactive computing platforms as an emerging form of `digital infrastructure' situates them alongside other digital 'platforms' that have coalesced into infrastructures (e.g. WeChat \cite{plantin2019wechat}, Facebook \cite{plantin2018infrastructure, helmond2019facebook, helmond2015platformization}, and Google \cite{plantin2018infrastructure}). The prominence of these digital infrastructures in mediating contemporary life has led to the development of the platform governance field \cite{gorwa2019what}.\footnote{Similarly, the emergence of earlier information infrastructures led to the development of the internet governance field \cite{hofmann2017coordination} and information infrastructure studies \cite{bowker2000sorting}.} Platform governance researchers have explored how digital infrastructures attempt to exercise governance over their users, and how digital infrastructures themselves can be more effectively governed. Robert Gorwa, for example, has studied the governance of online content, particularly user-generated content on digital platforms \cite{gorwa2019platform}. As Gorwa argues, there is an increasing nexus between AI ethics discourse and platform governance discourse: algorithmic systems, particularly predictive ML systems, are core components of the governance regimes of digital infrastructures \cite{gorwa2020fairness}. Tarleton Gillespie, for example, critiques the positioning of ML tools as the solution to social media content moderation \cite{gillespie2020content}. ICPs advance this nexus, but in the reverse direction: as the platforms have developed from software tools for scientific computing to general purpose coordination hubs for ML practices they have begun to integrate affordances more commonly associated with digital platforms. Google Colab, for example, integrates directly into Google Drive---a widely used cloud storage and file sharing platform. We can interpret this integration as an effort to cultivate network effects \cite{belleflamme2018platforms}: if I care about sharing my notebook with others, then it makes sense that I will seek out the ICP that integrates directly with the file sharing system most of my colleagues use. But, to the extent that a notebook is `content', and the extent that this content may include ML models that have been shown to cause significant social harm, ICP operators have so far eluded responsibility for this content. For the field of AI ethics, then, the potential for ICP operators to exercise governance functions over ICP users may be worth further consideration.\looseness=-1


%% file: sections/7_limitations.tex
Conceptually, as Eric Baumer and Micki McGee \cite{baumer2019speaking} argue, topic modelling risks using a statistical model of a corpus to speak on behalf of a social group. This risk is compounded by the fact that the social group who generates content that enters a corpus (in this study, people who ask questions on Stack Exchange forums) may not be representative of the social group of interest to the study (here, ML practitioners). Relevantly, among the Stack Overflow user base, as of 2016, only 5.8\% of contributors were female \cite{ford2016paradise}. Additionally, while there are versions of Stack Overflow in multiple languages, only the English-language version has been used in this study. As such, future research will need to validate the extent to which the practices identified in this study are representative of ML practitioners.\looseness=-1

The focus of Stack Exchange questions also presents a fundamental limitation for studies of ML practitioners. Stack Exchange questions are points of trouble---they represent moments when a user has been unable to complete a task. As such, it may be the case that there are a range of practices that are not represented in the Stack Exchange corpus, simply because they are practices so familiar they do not necessitate asking any questions. Given the discussion on transparency and infrastructures in Section \ref{subsec:infra_relations}, this means Stack Exchange questions can only offer a partial account of infrastructural relationships. There are also limitations inherent in the pre-processing and model training process outlined in Section \ref{subsec:estimation}. In particular, stemming of words may have reduced the semantic depth of the topic model, as may have removal of code snippets from the corpus. The validation of topic models is an ongoing area of research \cite{maier2018applying, grimmer2022text}. As this is a preliminary study, no attempt has been made to externally validate the accuracy of the topics identified (e.g. through comparing the latent topics identified by the topic model to coded themes identified by expert human reviewers of the same corpus, as in \cite{mimno2011optimizing}). More broadly, the approach taken in this study will benefit from complementary qualitative studies to both validate and contextualise findings (e.g. ethnographic studies of practitioners in multiple social contexts \cite{forsythe1993engineering}).\looseness=-1

%% file: sections/8_conclusion.tex
Interactive computing platforms, such as Jupyter Notebook and Google Colab, are widely used by ML practitioners. In this paper, I conducted a topic model analysis of user-contributed questions on the Stack Exchange forums related to interactive computing platforms and ML. I found interactive computing platforms are used by ML practitioners in two categories of practices: in learning practices, particularly to support probing and reuse of others' code; and, in coordination practices, to help marshal the various infrastructures needed to enact ML tasks. I argued that these practices constitute an emerging infrastructural relationship between ML practitioners and interactive computing platforms, in which both the platforms and ML practices are co-evolving. I highlighted several consequences of this infrastructuralisation, in terms of configuring the space in which AI ethics operates and responds to, designing interventions in ML practices, making visible the operationalisation in code of social constructs, and the platform power of ICP operators. As the ML field advances, a pressing issue is therefore the relationship between the social context ICPs form part of and the characteristics of ML systems that are developed. Tracing these relations is critical for resisting the enclosure of AI ethics by a set of social arrangements that may themselves be contributing to the production and deployment of harmful ML systems.

%% file: sections/9_appendix.tex
\section{List of tags used in query of the Stack Exchange Data Dump}
\label{subsec:dumptags}

\textbf{Interactive computing platform tags:} colab, google-colaboratory, ipython, ipython-notebook, ipywidgets, jupyter, jupyter-lab, jupyter-notebook, jupyterhub, pyspark.\\ \\
\noindent \textbf{Machine Learning tags:} artificial-intelligence, backpropagation, caffe, classification, cnn, computer-vision, conv-neural-network, convolutional-neural-network, deep-learning, feature-selection, \\image-processing, keras, lstm, machine-learning, machine-learning-model, neural-network, neural-networks, nlp, nltk, opencv, optimization, predictive-modelling, predictive-models, pytorch, random-forest, regression, scikit-learn, spacy, stanford-nlp, svm, tensorflow, tensorflow2.0.

\section{Topic model visualisations}
\label{subsec:stmdetails}

\begin{figure}[h]
    \centering
   \includegraphics[width=\linewidth]{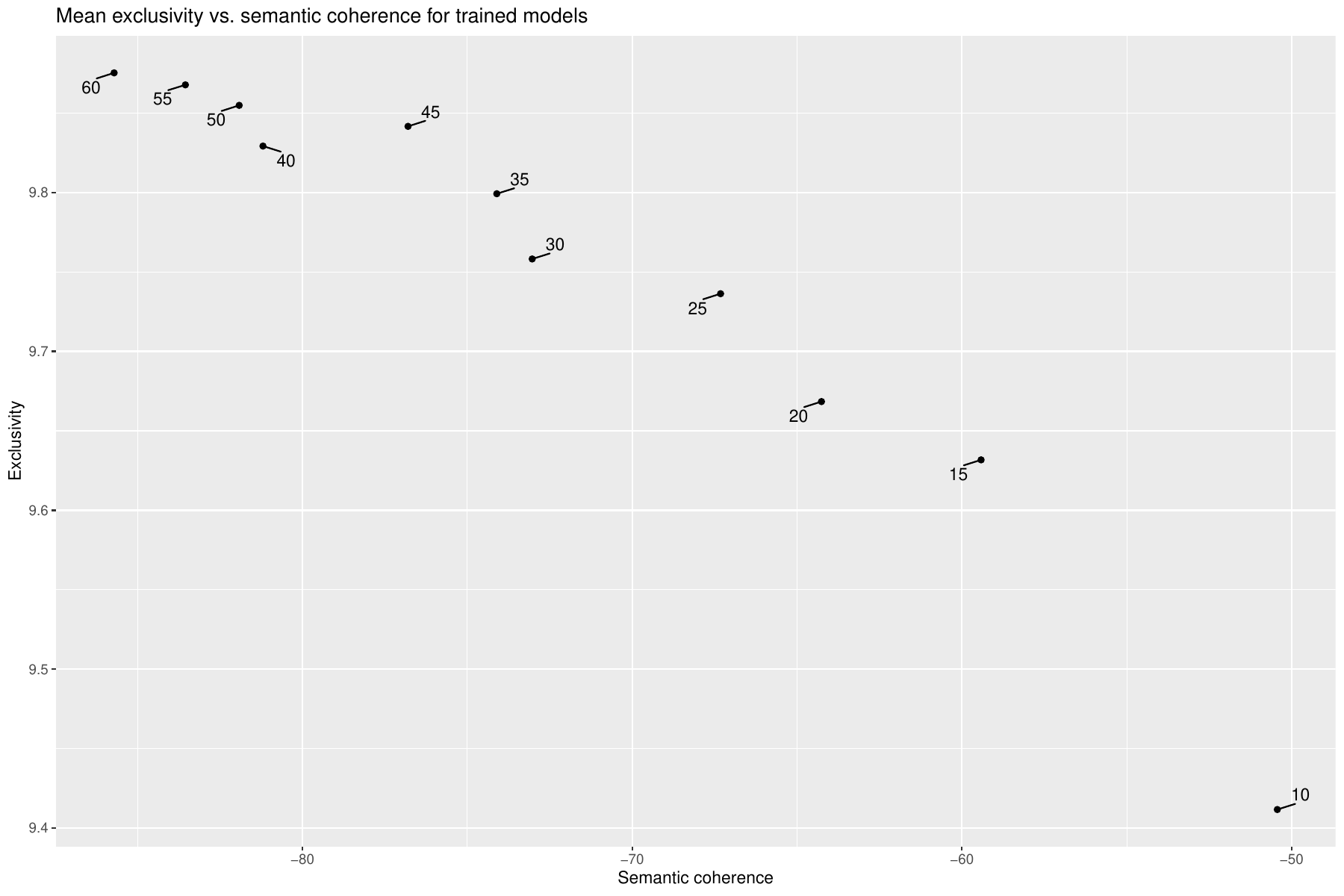}
    \caption{Exclusivity vs Semantic Coherence for a range of models trained on the Machine Learning in interactive computing platforms dataset.}
    \label{fig:excl_v_coherence}
\end{figure}

\begin{figure}[h]
    \centering
    \includegraphics[width=\linewidth]{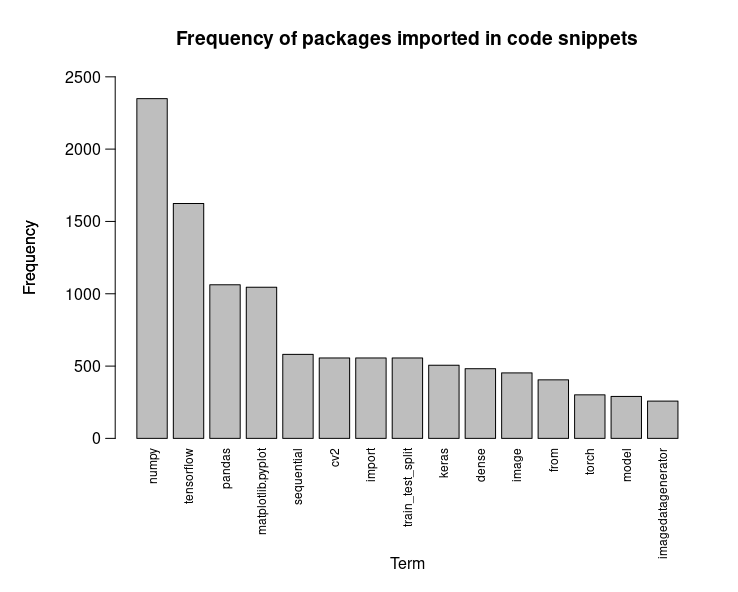}
    \caption{The 15 most frequently mentioned programming libraries imported in code snippets in questions about interactive computing platforms and ML on the Stack Exchange forums.}
    \label{fig:libraries}
\end{figure}

\onecolumn

\section{Representative questions by topic cluster}
\label{subsec:stmdetails2}

\begin{figure}[h]
    \centering
    \begin{minipage}{0.3\textwidth}
        \centering
        \includegraphics[width=\linewidth]{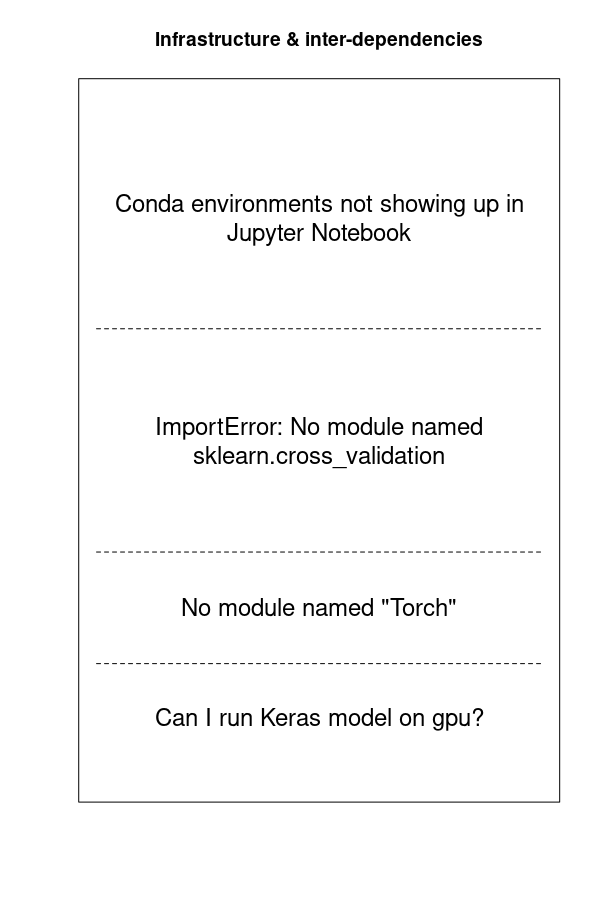}
        \caption{Most viewed questions: infra. \& inter-dependencies cluster.}
        \label{fig:cluster_quotes_1}    
    \end{minipage}%
    \hfill
    \begin{minipage}{0.3\textwidth}
        \centering
        \includegraphics[width=\linewidth]{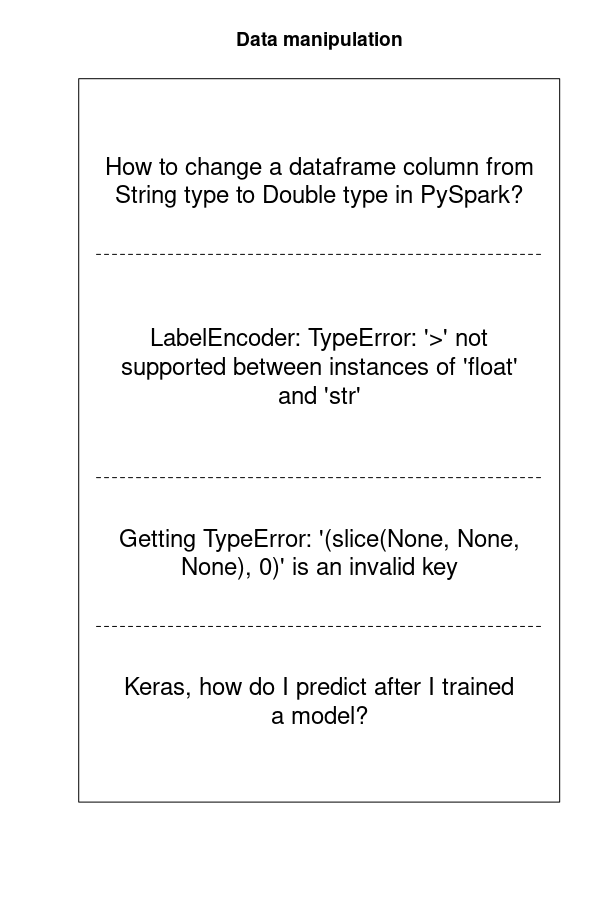}
        \caption{Most viewed questions: data manipulation cluster.}
        \label{fig:cluster_quotes_2}      
    \end{minipage}%
    \hfill
    \begin{minipage}{0.3\textwidth}
        \centering
        \includegraphics[width=\linewidth]{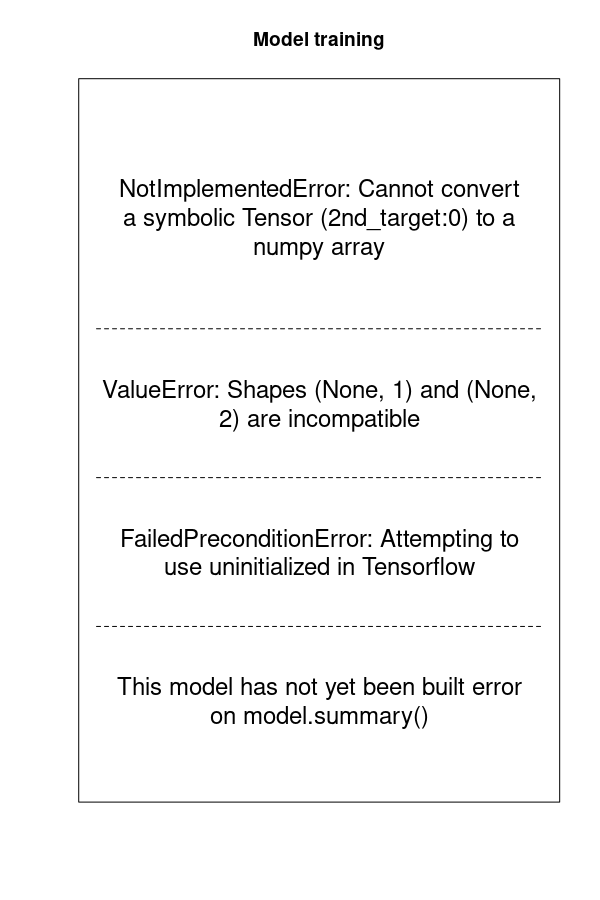}
        \caption{Most viewed questions: model training cluster.}
        \label{fig:cluster_quotes_3}        
    \end{minipage}
\end{figure}

%% file: main.bbl

\begin{thebibliography}{150}


\ifx \showCODEN    \undefined \def \showCODEN     #1{\unskip}     \fi
\ifx \showDOI      \undefined \def \showDOI       #1{#1}\fi
\ifx \showISBNx    \undefined \def \showISBNx     #1{\unskip}     \fi
\ifx \showISBNxiii \undefined \def \showISBNxiii  #1{\unskip}     \fi
\ifx \showISSN     \undefined \def \showISSN      #1{\unskip}     \fi
\ifx \showLCCN     \undefined \def \showLCCN      #1{\unskip}     \fi
\ifx \shownote     \undefined \def \shownote      #1{#1}          \fi
\ifx \showarticletitle \undefined \def \showarticletitle #1{#1}   \fi
\ifx \showURL      \undefined \def \showURL       {\relax}        \fi
\providecommand\bibfield[2]{#2}
\providecommand\bibinfo[2]{#2}
\providecommand\natexlab[1]{#1}
\providecommand\showeprint[2][]{arXiv:#2}

\bibitem[Abdalkareem et~al\mbox{.}(2017)]%
        {abdalkareem2017code}
\bibfield{author}{\bibinfo{person}{Rabe Abdalkareem}, \bibinfo{person}{Emad
  Shihab}, {and} \bibinfo{person}{Juergen Rilling}.}
  \bibinfo{year}{2017}\natexlab{}.
\newblock \showarticletitle{On Code Reuse from {{StackOverflow}}: {{An}}
  Exploratory Study on {{Android}} Apps}.
\newblock \bibinfo{journal}{\emph{Information and Software Technology}}
  \bibinfo{volume}{88} (\bibinfo{date}{Aug.} \bibinfo{year}{2017}).
\newblock


\bibitem[Ahmad et~al\mbox{.}(2018)]%
        {ahmad2018survey}
\bibfield{author}{\bibinfo{person}{Arshad Ahmad}, \bibinfo{person}{Chong Feng},
  \bibinfo{person}{Shi Ge}, {and} \bibinfo{person}{Abdallah Yousif}.}
  \bibinfo{year}{2018}\natexlab{}.
\newblock \showarticletitle{A Survey on Mining Stack Overflow: Question and
  Answering ({{Q}}\&{{A}}) Community}.
\newblock \bibinfo{journal}{\emph{Data Technologies and Applications}}
  \bibinfo{volume}{52}, \bibinfo{number}{2} (\bibinfo{date}{Jan.}
  \bibinfo{year}{2018}).
\newblock


\bibitem[Alshangiti et~al\mbox{.}(2019)]%
        {alshangiti2019why}
\bibfield{author}{\bibinfo{person}{Moayad Alshangiti}, \bibinfo{person}{Hitesh
  Sapkota}, \bibinfo{person}{Pradeep~K. Murukannaiah}, \bibinfo{person}{Xumin
  Liu}, {and} \bibinfo{person}{Qi Yu}.} \bibinfo{year}{2019}\natexlab{}.
\newblock \showarticletitle{Why Is {{Developing Machine Learning Applications
  Challenging}}? {{A Study}} on {{Stack Overflow Posts}}}. In
  \bibinfo{booktitle}{\emph{2019 {{ACM}}/{{IEEE International Symposium}} on
  {{Empirical Software Engineering}} and {{Measurement}} ({{ESEM}})}}.
  \bibinfo{publisher}{{IEEE}}, \bibinfo{address}{{Porto de Galinhas, Recife,
  Brazil}}.
\newblock


\bibitem[Amoore(2019)]%
        {amoore2019doubt}
\bibfield{author}{\bibinfo{person}{Louise Amoore}.}
  \bibinfo{year}{2019}\natexlab{}.
\newblock \showarticletitle{Doubt and the {{Algorithm}}: {{On}} the {{Partial
  Accounts}} of {{Machine Learning}}}.
\newblock \bibinfo{journal}{\emph{Theory, Culture \& Society}}
  \bibinfo{volume}{36}, \bibinfo{number}{6} (\bibinfo{date}{Nov.}
  \bibinfo{year}{2019}).
\newblock


\bibitem[An et~al\mbox{.}(2017)]%
        {an2017stack}
\bibfield{author}{\bibinfo{person}{Le An}, \bibinfo{person}{Ons Mlouki},
  \bibinfo{person}{Foutse Khomh}, {and} \bibinfo{person}{Giuliano Antoniol}.}
  \bibinfo{year}{2017}\natexlab{}.
\newblock \showarticletitle{Stack {{Overflow}}: {{A}} Code Laundering
  Platform?}. In \bibinfo{booktitle}{\emph{2017 {{IEEE}} 24th {{International
  Conference}} on {{Software Analysis}}, {{Evolution}} and {{Reengineering}}
  ({{SANER}})}}.
\newblock


\bibitem[Anderson et~al\mbox{.}(2012)]%
        {anderson2012discovering}
\bibfield{author}{\bibinfo{person}{Ashton Anderson}, \bibinfo{person}{Daniel
  Huttenlocher}, \bibinfo{person}{Jon Kleinberg}, {and} \bibinfo{person}{Jure
  Leskovec}.} \bibinfo{year}{2012}\natexlab{}.
\newblock \showarticletitle{Discovering Value from Community Activity on
  Focused Question Answering Sites: A Case Study of Stack Overflow}. In
  \bibinfo{booktitle}{\emph{Proceedings of the 18th {{ACM SIGKDD}}
  International Conference on {{Knowledge}} Discovery and Data Mining}}.
  \bibinfo{publisher}{{ACM}}, \bibinfo{address}{{Beijing China}}.
\newblock


\bibitem[Ashmore et~al\mbox{.}(2022)]%
        {ashmore2022assuring}
\bibfield{author}{\bibinfo{person}{Rob Ashmore}, \bibinfo{person}{Radu
  Calinescu}, {and} \bibinfo{person}{Colin Paterson}.}
  \bibinfo{year}{2022}\natexlab{}.
\newblock \showarticletitle{Assuring the {{Machine Learning Lifecycle}}:
  {{Desiderata}}, {{Methods}}, and {{Challenges}}}.
\newblock \bibinfo{journal}{\emph{Comput. Surveys}} \bibinfo{volume}{54},
  \bibinfo{number}{5} (\bibinfo{date}{June} \bibinfo{year}{2022}).
\newblock


\bibitem[Baltes and Diehl(2019)]%
        {baltes2019usage}
\bibfield{author}{\bibinfo{person}{Sebastian Baltes} {and}
  \bibinfo{person}{Stephan Diehl}.} \bibinfo{year}{2019}\natexlab{}.
\newblock \showarticletitle{Usage and Attribution of {{Stack Overflow}} Code
  Snippets in {{GitHub}} Projects}.
\newblock \bibinfo{journal}{\emph{Empirical Software Engineering}}
  \bibinfo{volume}{24}, \bibinfo{number}{3} (\bibinfo{date}{June}
  \bibinfo{year}{2019}).
\newblock


\bibitem[Barry et~al\mbox{.}(2020)]%
        {barry2020ethics}
\bibfield{author}{\bibinfo{person}{Marguerite Barry}, \bibinfo{person}{Aphra
  Kerr}, {and} \bibinfo{person}{Oliver Smith}.}
  \bibinfo{year}{2020}\natexlab{}.
\newblock \showarticletitle{Ethics on the {{Ground}}: {{From Principles}} to
  {{Practice}}}. In \bibinfo{booktitle}{\emph{Proceedings of the 2020
  {{Conference}} on {{Fairness}}, {{Accountability}}, and {{Transparency}}}}.
  \bibinfo{publisher}{{Association for Computing Machinery}}.
\newblock


\bibitem[Barua et~al\mbox{.}(2014)]%
        {barua2014what}
\bibfield{author}{\bibinfo{person}{Anton Barua}, \bibinfo{person}{Stephen~W.
  Thomas}, {and} \bibinfo{person}{Ahmed~E. Hassan}.}
  \bibinfo{year}{2014}\natexlab{}.
\newblock \showarticletitle{What Are Developers Talking about? {{An}} Analysis
  of Topics and Trends in {{Stack Overflow}}}.
\newblock \bibinfo{journal}{\emph{Empirical Software Engineering}}
  \bibinfo{volume}{19}, \bibinfo{number}{3} (\bibinfo{date}{June}
  \bibinfo{year}{2014}).
\newblock


\bibitem[Baumer and McGee(2019)]%
        {baumer2019speaking}
\bibfield{author}{\bibinfo{person}{Eric P.~S. Baumer} {and}
  \bibinfo{person}{Micki McGee}.} \bibinfo{year}{2019}\natexlab{}.
\newblock \showarticletitle{Speaking on {{Behalf}} of: {{Representation}},
  {{Delegation}}, and {{Authority}} in {{Computational Text Analysis}}}. In
  \bibinfo{booktitle}{\emph{Proceedings of the 2019 {{AAAI}}/{{ACM Conference}}
  on {{AI}}, {{Ethics}}, and {{Society}}}}. \bibinfo{publisher}{{ACM}},
  \bibinfo{address}{{Honolulu HI USA}}.
\newblock


\bibitem[Baxter et~al\mbox{.}(2020)]%
        {baxter2020bridging}
\bibfield{author}{\bibinfo{person}{Kathy Baxter}, \bibinfo{person}{Yoav
  Schlesinger}, \bibinfo{person}{Sarah Aerni}, \bibinfo{person}{Lewis Baker},
  \bibinfo{person}{Julie Dawson}, \bibinfo{person}{Krishnaram Kenthapadi},
  \bibinfo{person}{Isabel Kloumann}, {and} \bibinfo{person}{Hanna Wallach}.}
  \bibinfo{year}{2020}\natexlab{}.
\newblock \showarticletitle{Bridging the {{Gap}} from {{AI Ethics Research}} to
  {{Practice}}}.
\newblock \bibinfo{journal}{\emph{Proceedings of the 2020 Conference on
  Fairness, Accountability, and Transparency}} (\bibinfo{year}{2020}).
\newblock


\bibitem[Beg et~al\mbox{.}(2021)]%
        {beg2021using}
\bibfield{author}{\bibinfo{person}{Marijan Beg}, \bibinfo{person}{Juliette
  Taka}, \bibinfo{person}{Thomas Kluyver}, \bibinfo{person}{Alexander
  Konovalov}, \bibinfo{person}{Min {Ragan-Kelley}}, \bibinfo{person}{Nicolas~M.
  Thi{\'e}ry}, {and} \bibinfo{person}{Hans Fangohr}.}
  \bibinfo{year}{2021}\natexlab{}.
\newblock \showarticletitle{Using {{Jupyter}} for Reproducible Scientific
  Workflows}.
\newblock \bibinfo{journal}{\emph{Computing in Science \& Engineering}}
  \bibinfo{volume}{23}, \bibinfo{number}{2} (\bibinfo{year}{2021}).
\newblock


\bibitem[Belleflamme(2018)]%
        {belleflamme2018platforms}
\bibfield{author}{\bibinfo{person}{Paul Belleflamme}.}
  \bibinfo{year}{2018}\natexlab{}.
\newblock \showarticletitle{Platforms and Network Effects}.
\newblock In \bibinfo{booktitle}{\emph{Handbook of {{Game Theory}} and
  {{Industrial Organization}}, {{Volume II}}}},
  \bibfield{editor}{\bibinfo{person}{Luis Corch{\'o}n} {and}
  \bibinfo{person}{Marco Marini}} (Eds.). \bibinfo{publisher}{{Edward Elgar
  Publishing}}.
\newblock


\bibitem[Bessen et~al\mbox{.}(2022)]%
        {bessen2022cost}
\bibfield{author}{\bibinfo{person}{James Bessen},
  \bibinfo{person}{Stephen~Michael Impink}, {and} \bibinfo{person}{Robert
  Seamans}.} \bibinfo{year}{2022}\natexlab{}.
\newblock \showarticletitle{The {{Cost}} of {{Ethical AI Development}} for {{AI
  Startups}}}. In \bibinfo{booktitle}{\emph{Proceedings of the 2022
  {{AAAI}}/{{ACM Conference}} on {{AI}}, {{Ethics}}, and {{Society}}}}.
  \bibinfo{publisher}{{ACM}}, \bibinfo{address}{{Oxford United Kingdom}}.
\newblock


\bibitem[Beutel et~al\mbox{.}(2019)]%
        {beutel2019putting}
\bibfield{author}{\bibinfo{person}{Alex Beutel}, \bibinfo{person}{Jilin Chen},
  \bibinfo{person}{Tulsee Doshi}, \bibinfo{person}{Hai Qian},
  \bibinfo{person}{Allison Woodruff}, \bibinfo{person}{Christine Luu},
  \bibinfo{person}{Pierre Kreitmann}, \bibinfo{person}{Jonathan Bischof}, {and}
  \bibinfo{person}{Ed~H. Chi}.} \bibinfo{year}{2019}\natexlab{}.
\newblock \showarticletitle{Putting {{Fairness Principles}} into {{Practice}}:
  {{Challenges}}, {{Metrics}}, and {{Improvements}}}. In
  \bibinfo{booktitle}{\emph{Proceedings of the 2019 {{AAAI}}/{{ACM Conference}}
  on {{AI}}, {{Ethics}}, and {{Society}}}} \emph{(\bibinfo{series}{{{AIES}}
  '19})}. \bibinfo{publisher}{{Association for Computing Machinery}},
  \bibinfo{address}{{New York, NY, USA}}.
\newblock


\bibitem[Bogen et~al\mbox{.}(2020)]%
        {bogen2020awareness}
\bibfield{author}{\bibinfo{person}{Miranda Bogen}, \bibinfo{person}{Aaron
  Rieke}, {and} \bibinfo{person}{Shazeda Ahmed}.}
  \bibinfo{year}{2020}\natexlab{}.
\newblock \showarticletitle{Awareness in {{Practice}}: {{Tensions}} in
  {{Access}} to {{Sensitive Attribute Data}} for {{Antidiscrimination}}}. In
  \bibinfo{booktitle}{\emph{Proceedings of the 2020 {{Conference}} on
  {{Fairness}}, {{Accountability}}, and {{Transparency}}}}.
  \bibinfo{publisher}{{Association for Computing Machinery}}.
\newblock


\bibitem[Bourdieu(2020)]%
        {bourdieu2020outline}
\bibfield{author}{\bibinfo{person}{Pierre Bourdieu}.}
  \bibinfo{year}{2020}\natexlab{}.
\newblock \showarticletitle{Outline of a {{Theory}} of {{Practice}}}.
\newblock In \bibinfo{booktitle}{\emph{The New Social Theory Reader}}.
  \bibinfo{publisher}{{Routledge}}.
\newblock


\bibitem[Bowker et~al\mbox{.}(2009)]%
        {bowker2009information}
\bibfield{author}{\bibinfo{person}{Geoffrey~C. Bowker}, \bibinfo{person}{Karen
  Baker}, \bibinfo{person}{Florence Millerand}, {and} \bibinfo{person}{David
  Ribes}.} \bibinfo{year}{2009}\natexlab{}.
\newblock \showarticletitle{Toward {{Information Infrastructure Studies}}:
  {{Ways}} of {{Knowing}} in a {{Networked Environment}}}.
\newblock In \bibinfo{booktitle}{\emph{International {{Handbook}} of {{Internet
  Research}}}}, \bibfield{editor}{\bibinfo{person}{Jeremy Hunsinger},
  \bibinfo{person}{Lisbeth Klastrup}, {and} \bibinfo{person}{Matthew Allen}}
  (Eds.). \bibinfo{publisher}{{Springer Netherlands}},
  \bibinfo{address}{{Dordrecht}}.
\newblock


\bibitem[Bowker and Star(2000)]%
        {bowker2000sorting}
\bibfield{author}{\bibinfo{person}{Geoffrey~C. Bowker} {and}
  \bibinfo{person}{Susan~Leigh Star}.} \bibinfo{year}{2000}\natexlab{}.
\newblock \bibinfo{booktitle}{\emph{Sorting Things out: Classification and Its
  Consequences}}.
\newblock \bibinfo{publisher}{{MIT Press}}.
\newblock


\bibitem[Brinkmann(2021)]%
        {brinkmann2021jupyter}
\bibfield{author}{\bibinfo{person}{Demetrios Brinkmann}.}
  \bibinfo{year}{2021}\natexlab{}.
\newblock \bibinfo{title}{Jupyter {{Notebooks In Production}}?}
\newblock
\newblock


\bibitem[Brock(2018)]%
        {brock2018critical}
\bibfield{author}{\bibinfo{person}{Andr{\'e} Brock}.}
  \bibinfo{year}{2018}\natexlab{}.
\newblock \showarticletitle{Critical Technocultural Discourse Analysis}.
\newblock \bibinfo{journal}{\emph{New Media and Society}} \bibinfo{volume}{20},
  \bibinfo{number}{3} (\bibinfo{year}{2018}).
\newblock
\showISSN{14617315}


\bibitem[Brookes and McEnery(2019)]%
        {brookes2019utility}
\bibfield{author}{\bibinfo{person}{Gavin Brookes} {and} \bibinfo{person}{Tony
  McEnery}.} \bibinfo{year}{2019}\natexlab{}.
\newblock \showarticletitle{The Utility of Topic Modelling for Discourse
  Studies: {{A}} Critical Evaluation}.
\newblock \bibinfo{journal}{\emph{Discourse Studies}} \bibinfo{volume}{21},
  \bibinfo{number}{1} (\bibinfo{date}{Feb.} \bibinfo{year}{2019}).
\newblock


\bibitem[Brown et~al\mbox{.}(2021)]%
        {brown2021reproducing}
\bibfield{author}{\bibinfo{person}{Duncan~A. Brown}, \bibinfo{person}{Karan
  Vahi}, \bibinfo{person}{Michela Taufer}, \bibinfo{person}{Von Welch}, {and}
  \bibinfo{person}{Ewa Deelman}.} \bibinfo{year}{2021}\natexlab{}.
\newblock \showarticletitle{Reproducing {{GW150914}}: {{The First Observation}}
  of {{Gravitational Waves From}} a {{Binary Black Hole Merger}}}.
\newblock \bibinfo{journal}{\emph{Computing in Science Engineering}}
  \bibinfo{volume}{23}, \bibinfo{number}{2} (\bibinfo{date}{March}
  \bibinfo{year}{2021}).
\newblock


\bibitem[Buolamwini and Gebru(2018)]%
        {buolamwini2018gender}
\bibfield{author}{\bibinfo{person}{Joy Buolamwini} {and}
  \bibinfo{person}{Timnit Gebru}.} \bibinfo{year}{2018}\natexlab{}.
\newblock \showarticletitle{Gender shades: Intersectional accuracy disparities
  in commercial gender classification}. In \bibinfo{booktitle}{\emph{Conference
  on fairness, accountability and transparency}}. PMLR,
  \bibinfo{pages}{77--91}.
\newblock


\bibitem[Burrell(2016)]%
        {burrell2016how}
\bibfield{author}{\bibinfo{person}{Jenna Burrell}.}
  \bibinfo{year}{2016}\natexlab{}.
\newblock \showarticletitle{How the Machine `thinks': {{Understanding}} Opacity
  in Machine Learning Algorithms}.
\newblock \bibinfo{journal}{\emph{Big Data and Society}} \bibinfo{volume}{3},
  \bibinfo{number}{1} (\bibinfo{year}{2016}).
\newblock


\bibitem[Chattopadhyay et~al\mbox{.}(2020)]%
        {chattopadhyay2020what}
\bibfield{author}{\bibinfo{person}{Souti Chattopadhyay},
  \bibinfo{person}{Ishita Prasad}, \bibinfo{person}{Austin~Z. Henley},
  \bibinfo{person}{Anita Sarma}, {and} \bibinfo{person}{Titus Barik}.}
  \bibinfo{year}{2020}\natexlab{}.
\newblock \showarticletitle{What's {{Wrong}} with {{Computational Notebooks}}?
  {{Pain Points}}, {{Needs}}, and {{Design Opportunities}}}. In
  \bibinfo{booktitle}{\emph{Proceedings of the 2020 {{CHI Conference}} on
  {{Human Factors}} in {{Computing Systems}}}}. \bibinfo{publisher}{{ACM}},
  \bibinfo{address}{{Honolulu HI USA}}.
\newblock


\bibitem[Christin(2017)]%
        {christin2017algorithms}
\bibfield{author}{\bibinfo{person}{Ang{\`e}le Christin}.}
  \bibinfo{year}{2017}\natexlab{}.
\newblock \showarticletitle{Algorithms in Practice: {{Comparing}} Web
  Journalism and Criminal Justice}.
\newblock \bibinfo{journal}{\emph{Big Data \& Society}} \bibinfo{volume}{4},
  \bibinfo{number}{2} (\bibinfo{date}{Dec.} \bibinfo{year}{2017}).
\newblock


\bibitem[Cooper et~al\mbox{.}(2022)]%
        {cooper2022accountability}
\bibfield{author}{\bibinfo{person}{A.~Feder Cooper}, \bibinfo{person}{Emanuel
  Moss}, \bibinfo{person}{Benjamin Laufer}, {and} \bibinfo{person}{Helen
  Nissenbaum}.} \bibinfo{year}{2022}\natexlab{}.
\newblock \showarticletitle{Accountability in an {{Algorithmic Society}}:
  {{Relationality}}, {{Responsibility}}, and {{Robustness}} in {{Machine
  Learning}}}. In \bibinfo{booktitle}{\emph{2022 {{ACM Conference}} on
  {{Fairness}}, {{Accountability}}, and {{Transparency}}}}.
  \bibinfo{publisher}{{ACM}}, \bibinfo{address}{{Seoul Republic of Korea}}.
\newblock


\bibitem[Davis(2020)]%
        {davis2020artifacts}
\bibfield{author}{\bibinfo{person}{Jenny~L Davis}.}
  \bibinfo{year}{2020}\natexlab{}.
\newblock \bibinfo{booktitle}{\emph{How artifacts afford: The power and
  politics of everyday things}}.
\newblock \bibinfo{publisher}{MIT Press}.
\newblock


\bibitem[Davis and Chouinard(2016)]%
        {davis2016theorizing}
\bibfield{author}{\bibinfo{person}{Jenny~L. Davis} {and}
  \bibinfo{person}{James~B. Chouinard}.} \bibinfo{year}{2016}\natexlab{}.
\newblock \showarticletitle{Theorizing {{Affordances}}: {{From Request}} to
  {{Refuse}}}.
\newblock \bibinfo{journal}{\emph{Bulletin of Science, Technology \& Society}}
  \bibinfo{volume}{36}, \bibinfo{number}{4} (\bibinfo{year}{2016}).
\newblock


\bibitem[Denton et~al\mbox{.}(2021)]%
        {denton2021genealogy}
\bibfield{author}{\bibinfo{person}{Emily Denton}, \bibinfo{person}{Alex Hanna},
  \bibinfo{person}{Razvan Amironesei}, \bibinfo{person}{Andrew Smart}, {and}
  \bibinfo{person}{Hilary Nicole}.} \bibinfo{year}{2021}\natexlab{}.
\newblock \showarticletitle{On the Genealogy of Machine Learning Datasets:
  {{A}} Critical History of {{ImageNet}}}.
\newblock \bibinfo{journal}{\emph{Big Data \& Society}} \bibinfo{volume}{8},
  \bibinfo{number}{2} (\bibinfo{date}{July} \bibinfo{year}{2021}).
\newblock


\bibitem[Deshpande and Sharp(2022)]%
        {deshpande2022responsible}
\bibfield{author}{\bibinfo{person}{Advait Deshpande} {and}
  \bibinfo{person}{Helen Sharp}.} \bibinfo{year}{2022}\natexlab{}.
\newblock \showarticletitle{Responsible {{AI Systems}}: {{Who}} Are the
  {{Stakeholders}}?}. In \bibinfo{booktitle}{\emph{Proceedings of the 2022
  {{AAAI}}/{{ACM Conference}} on {{AI}}, {{Ethics}}, and {{Society}}}}.
  \bibinfo{publisher}{{ACM}}, \bibinfo{address}{{Oxford United Kingdom}}.
\newblock


\bibitem[DiMaggio et~al\mbox{.}(2013)]%
        {dimaggio2013exploiting}
\bibfield{author}{\bibinfo{person}{Paul DiMaggio}, \bibinfo{person}{Manish
  Nag}, {and} \bibinfo{person}{David Blei}.} \bibinfo{year}{2013}\natexlab{}.
\newblock \showarticletitle{Exploiting Affinities between Topic Modeling and
  the Sociological Perspective on Culture: {{Application}} to Newspaper
  Coverage of {{U}}.{{S}}. Government Arts Funding}.
\newblock \bibinfo{journal}{\emph{Poetics}} \bibinfo{volume}{41},
  \bibinfo{number}{6} (\bibinfo{date}{Dec.} \bibinfo{year}{2013}).
\newblock
\showISSN{0304-422X}


\bibitem[Dourish(2016)]%
        {dourish2016algorithms}
\bibfield{author}{\bibinfo{person}{Paul Dourish}.}
  \bibinfo{year}{2016}\natexlab{}.
\newblock \showarticletitle{Algorithms and Their Others: {{Algorithmic}}
  Culture in Context}.
\newblock \bibinfo{journal}{\emph{Big Data and Society}} \bibinfo{volume}{3},
  \bibinfo{number}{2} (\bibinfo{year}{2016}).
\newblock


\bibitem[Emery(1993)]%
        {emery1993characteristics}
\bibfield{author}{\bibinfo{person}{Fred Emery}.}
  \bibinfo{year}{1993}\natexlab{}.
\newblock \showarticletitle{Characteristics of {{Socio-Technical Systems}}}.
\newblock In \bibinfo{booktitle}{\emph{The {{Social Engagement}} of {{Social
  Science}}, {{Volume}} 2}}, \bibfield{editor}{\bibinfo{person}{Eric Trist},
  \bibinfo{person}{Hugh Murray}, {and} \bibinfo{person}{Beulah Trist}} (Eds.).
  \bibinfo{publisher}{{University of Pennsylvania Press}},
  \bibinfo{address}{{Philadelphia}}.
\newblock


\bibitem[Fish and Stark(2021)]%
        {fish2021reflexive}
\bibfield{author}{\bibinfo{person}{Benjamin Fish} {and} \bibinfo{person}{Luke
  Stark}.} \bibinfo{year}{2021}\natexlab{}.
\newblock \showarticletitle{Reflexive {{Design}} for {{Fairness}} and {{Other
  Human Values}} in {{Formal Models}}}. In
  \bibinfo{booktitle}{\emph{Proceedings of the 2021 {{AAAI}}/{{ACM Conference}}
  on {{AI}}, {{Ethics}}, and {{Society}}}}. \bibinfo{publisher}{{ACM}},
  \bibinfo{address}{{Virtual Event USA}}.
\newblock


\bibitem[Ford et~al\mbox{.}(2016)]%
        {ford2016paradise}
\bibfield{author}{\bibinfo{person}{Denae Ford}, \bibinfo{person}{Justin Smith},
  \bibinfo{person}{Philip~J. Guo}, {and} \bibinfo{person}{Chris Parnin}.}
  \bibinfo{year}{2016}\natexlab{}.
\newblock \showarticletitle{Paradise Unplugged: Identifying Barriers for Female
  Participation on Stack Overflow}. In \bibinfo{booktitle}{\emph{Proceedings of
  the 2016 24th {{ACM SIGSOFT International Symposium}} on {{Foundations}} of
  {{Software Engineering}}}}. \bibinfo{publisher}{{ACM}},
  \bibinfo{address}{{Seattle WA USA}}.
\newblock


\bibitem[Forsythe(1993)]%
        {forsythe1993engineering}
\bibfield{author}{\bibinfo{person}{Diana~E. Forsythe}.}
  \bibinfo{year}{1993}\natexlab{}.
\newblock \showarticletitle{Engineering {{Knowledge}}: {{The Construction}} of
  {{Knowledge}} in {{Artificial Intelligence}}}.
\newblock \bibinfo{journal}{\emph{Social Studies of Science}}
  \bibinfo{volume}{23}, \bibinfo{number}{3} (\bibinfo{date}{Aug.}
  \bibinfo{year}{1993}).
\newblock


\bibitem[Gebru et~al\mbox{.}(2021)]%
        {gebru2021datasheets}
\bibfield{author}{\bibinfo{person}{Timnit Gebru}, \bibinfo{person}{Jamie
  Morgenstern}, \bibinfo{person}{Briana Vecchione},
  \bibinfo{person}{Jennifer~Wortman Vaughan}, \bibinfo{person}{Hanna Wallach},
  \bibinfo{person}{Hal~Daum{\'e} Iii}, {and} \bibinfo{person}{Kate Crawford}.}
  \bibinfo{year}{2021}\natexlab{}.
\newblock \showarticletitle{Datasheets for Datasets}.
\newblock \bibinfo{journal}{\emph{Commun. ACM}} \bibinfo{volume}{64},
  \bibinfo{number}{12} (\bibinfo{date}{Dec.} \bibinfo{year}{2021}).
\newblock
\showISSN{0001-0782, 1557-7317}


\bibitem[Gillespie(2020)]%
        {gillespie2020content}
\bibfield{author}{\bibinfo{person}{Tarleton Gillespie}.}
  \bibinfo{year}{2020}\natexlab{}.
\newblock \showarticletitle{Content Moderation, {{AI}}, and the Question of
  Scale}.
\newblock \bibinfo{journal}{\emph{Big Data and Society}}  \bibinfo{volume}{7}
  (\bibinfo{year}{2020}).
\newblock


\bibitem[Gorwa(2019a)]%
        {gorwa2019platform}
\bibfield{author}{\bibinfo{person}{Robert Gorwa}.}
  \bibinfo{year}{2019}\natexlab{a}.
\newblock \showarticletitle{The Platform Governance Triangle: Conceptualising
  the Informal Regulation of Online Content}.
\newblock \bibinfo{journal}{\emph{Internet Policy Review}} \bibinfo{volume}{8},
  \bibinfo{number}{2} (\bibinfo{date}{June} \bibinfo{year}{2019}).
\newblock


\bibitem[Gorwa(2019b)]%
        {gorwa2019what}
\bibfield{author}{\bibinfo{person}{Robert Gorwa}.}
  \bibinfo{year}{2019}\natexlab{b}.
\newblock \showarticletitle{What Is Platform Governance?}
\newblock \bibinfo{journal}{\emph{Information, Communication \& Society}}
  \bibinfo{volume}{22}, \bibinfo{number}{6} (\bibinfo{date}{May}
  \bibinfo{year}{2019}).
\newblock


\bibitem[Gorwa(2020)]%
        {gorwa2020fairness}
\bibfield{author}{\bibinfo{person}{Robert Gorwa}.}
  \bibinfo{year}{2020}\natexlab{}.
\newblock \showarticletitle{Towards Fairness, Accountability, and Transparency
  in Platform Govvernance}.
\newblock \bibinfo{journal}{\emph{AoIR Selected Papers of Internet Research}}
  (\bibinfo{date}{Feb.} \bibinfo{year}{2020}).
\newblock


\bibitem[Granger and P{\'e}rez(2021)]%
        {granger2021jupyter}
\bibfield{author}{\bibinfo{person}{Brian Granger} {and}
  \bibinfo{person}{Fernando P{\'e}rez}.} \bibinfo{year}{2021}\natexlab{}.
\newblock \bibinfo{booktitle}{\emph{Jupyter: {{Thinking}} and {{Storytelling}}
  with {{Code}} and {{Data}}}}.
\newblock \bibinfo{type}{Preprint}.
\newblock


\bibitem[Grimmer et~al\mbox{.}(2022)]%
        {grimmer2022text}
\bibfield{author}{\bibinfo{person}{Justin Grimmer}, \bibinfo{person}{{Roberts.
  Margaret E.}}, {and} \bibinfo{person}{{Stewart, Brandon M.}}}
  \bibinfo{year}{2022}\natexlab{}.
\newblock \bibinfo{booktitle}{\emph{Text as {{Data}}: {{A New Framework}} for
  {{Machine Learning}} and the {{Social Sciences}}}}.
\newblock \bibinfo{publisher}{{Princeton}}.
\newblock


\bibitem[Grimmer and Stewart(2013)]%
        {grimmer2013text}
\bibfield{author}{\bibinfo{person}{Justin Grimmer} {and}
  \bibinfo{person}{Brandon~M. Stewart}.} \bibinfo{year}{2013}\natexlab{}.
\newblock \showarticletitle{Text as {{Data}}: {{The Promise}} and {{Pitfalls}}
  of {{Automatic Content Analysis Methods}} for {{Political Texts}}}.
\newblock \bibinfo{journal}{\emph{Political Analysis}} \bibinfo{volume}{21},
  \bibinfo{number}{3} (\bibinfo{year}{2013}).
\newblock


\bibitem[Grus(2018)]%
        {grusdon2018i}
\bibfield{author}{\bibinfo{person}{Joel Grus}.}
  \bibinfo{year}{2018}\natexlab{}.
\newblock \bibinfo{title}{I Don't like Notebooks.: {{Jupyter Notebook}}
  Conference \& Training: {{JupyterCon}}}.
\newblock
\newblock
\urldef\tempurl%
\url{https://conferences.oreilly.com/jupyter/jup-ny/public/schedule/detail/68282.html}
\showURL{%
\tempurl}


\bibitem[Guribye(2015)]%
        {guribye2015artifacts}
\bibfield{author}{\bibinfo{person}{Frode Guribye}.}
  \bibinfo{year}{2015}\natexlab{}.
\newblock \showarticletitle{From {{Artifacts}} to {{Infrastructures}} in
  {{Studies}} of {{Learning Practices}}}.
\newblock \bibinfo{journal}{\emph{Mind, Culture, and Activity}}
  \bibinfo{volume}{22}, \bibinfo{number}{2} (\bibinfo{date}{April}
  \bibinfo{year}{2015}).
\newblock
\showISSN{1074-9039, 1532-7884}


\bibitem[Hardt et~al\mbox{.}(2021)]%
        {hardt2021amazon}
\bibfield{author}{\bibinfo{person}{Michaela Hardt}, \bibinfo{person}{Xiaoguang
  Chen}, \bibinfo{person}{Xiaoyi Cheng}, \bibinfo{person}{Michele Donini},
  \bibinfo{person}{Jason Gelman}, \bibinfo{person}{Satish Gollaprolu},
  \bibinfo{person}{John He}, \bibinfo{person}{Pedro Larroy},
  \bibinfo{person}{Xinyu Liu}, \bibinfo{person}{Nick McCarthy},
  \bibinfo{person}{Ashish Rathi}, \bibinfo{person}{Scott Rees},
  \bibinfo{person}{Ankit Siva}, \bibinfo{person}{ErhYuan Tsai},
  \bibinfo{person}{Keerthan Vasist}, \bibinfo{person}{Pinar Yilmaz},
  \bibinfo{person}{Muhammad~Bilal Zafar}, \bibinfo{person}{Sanjiv Das},
  \bibinfo{person}{Kevin Haas}, \bibinfo{person}{Tyler Hill}, {and}
  \bibinfo{person}{Krishnaram Kenthapadi}.} \bibinfo{year}{2021}\natexlab{}.
\newblock \showarticletitle{Amazon {{SageMaker Clarify}}: {{Machine Learning
  Bias Detection}} and {{Explainability}} in the {{Cloud}}}. In
  \bibinfo{booktitle}{\emph{Proceedings of the 27th {{ACM SIGKDD Conference}}
  on {{Knowledge Discovery}} \& {{Data Mining}}}}. \bibinfo{publisher}{{ACM}},
  \bibinfo{address}{{Virtual Event Singapore}}.
\newblock


\bibitem[Helmond(2015)]%
        {helmond2015platformization}
\bibfield{author}{\bibinfo{person}{Anne Helmond}.}
  \bibinfo{year}{2015}\natexlab{}.
\newblock \showarticletitle{The {{Platformization}} of the {{Web}}: {{Making
  Web Data Platform Ready}}}.
\newblock \bibinfo{journal}{\emph{Social Media and Society}}
  \bibinfo{volume}{1}, \bibinfo{number}{2} (\bibinfo{year}{2015}).
\newblock


\bibitem[Helmond et~al\mbox{.}(2019)]%
        {helmond2019facebook}
\bibfield{author}{\bibinfo{person}{Anne Helmond}, \bibinfo{person}{David~B.
  Nieborg}, {and} \bibinfo{person}{Fernando~N. {van der Vlist}}.}
  \bibinfo{year}{2019}\natexlab{}.
\newblock \showarticletitle{Facebook's Evolution: Development of a
  Platform-as-Infrastructure}.
\newblock \bibinfo{journal}{\emph{Internet Histories}} \bibinfo{volume}{3},
  \bibinfo{number}{2} (\bibinfo{date}{April} \bibinfo{year}{2019}).
\newblock


\bibitem[Hofmann et~al\mbox{.}(2017)]%
        {hofmann2017coordination}
\bibfield{author}{\bibinfo{person}{Jeanette Hofmann},
  \bibinfo{person}{Christian Katzenbach}, {and} \bibinfo{person}{Kirsten
  Gollatz}.} \bibinfo{year}{2017}\natexlab{}.
\newblock \showarticletitle{Between Coordination and Regulation: {{Finding}}
  the Governance in {{Internet}} Governance}.
\newblock \bibinfo{journal}{\emph{New Media \& Society}} \bibinfo{volume}{19},
  \bibinfo{number}{9} (\bibinfo{date}{Sept.} \bibinfo{year}{2017}).
\newblock


\bibitem[Holstein et~al\mbox{.}(2019)]%
        {holstein2019improving}
\bibfield{author}{\bibinfo{person}{Kenneth Holstein},
  \bibinfo{person}{Jennifer~Wortman Vaughan}, \bibinfo{person}{Hal Daum{\'e}},
  \bibinfo{person}{Miroslav Dud{\'i}k}, {and} \bibinfo{person}{Hanna Wallach}.}
  \bibinfo{year}{2019}\natexlab{}.
\newblock \showarticletitle{Improving {{Fairness}} in {{Machine Learning
  Systems}}: {{What Do Industry Practitioners Need}}?}. In
  \bibinfo{booktitle}{\emph{Proceedings of the 2019 {{CHI Conference}} on
  {{Human Factors}} in {{Computing Systems}}}}.
\newblock


\bibitem[Hopkins and Booth(2021)]%
        {hopkins2021machine}
\bibfield{author}{\bibinfo{person}{Aspen Hopkins} {and} \bibinfo{person}{Serena
  Booth}.} \bibinfo{year}{2021}\natexlab{}.
\newblock \showarticletitle{Machine {{Learning Practices Outside Big Tech}}:
  {{How Resource Constraints Challenge Responsible Development}}}. In
  \bibinfo{booktitle}{\emph{Proceedings of the 2021 {{AAAI}}/{{ACM Conference}}
  on {{AI}}, {{Ethics}}, and {{Society}}}}. \bibinfo{publisher}{{Association
  for Computing Machinery}}.
\newblock


\bibitem[Howard(2020)]%
        {howard2020creating}
\bibfield{author}{\bibinfo{person}{Jeremy Howard}.}
  \bibinfo{year}{2020}\natexlab{}.
\newblock \bibinfo{title}{Creating Delightful Libraries and Books with Nbdev
  and Fastdoc}.
\newblock
\newblock


\bibitem[Hughes(1987)]%
        {hughes1987evolution}
\bibfield{author}{\bibinfo{person}{Thomas~P. Hughes}.}
  \bibinfo{year}{1987}\natexlab{}.
\newblock \showarticletitle{The Evolution of Large Technological Systems}.
\newblock \bibinfo{journal}{\emph{The social construction of technological
  systems: New directions in the sociology and history of technology}}
  \bibinfo{volume}{82} (\bibinfo{year}{1987}).
\newblock


\bibitem[Ingram et~al\mbox{.}(2007)]%
        {ingram2007products}
\bibfield{author}{\bibinfo{person}{Jack Ingram}, \bibinfo{person}{Elizabeth
  Shove}, {and} \bibinfo{person}{Matthew Watson}.}
  \bibinfo{year}{2007}\natexlab{}.
\newblock \showarticletitle{Products and {{Practices}}: {{Selected Concepts}}
  from {{Science}} and {{Technology Studies}} and from {{Social Theories}} of
  {{Consumption}} and {{Practice}}}.
\newblock \bibinfo{journal}{\emph{Design Issues}} \bibinfo{volume}{23},
  \bibinfo{number}{2} (\bibinfo{year}{2007}).
\newblock


\bibitem[Isoaho et~al\mbox{.}(2021)]%
        {isoaho2021topic}
\bibfield{author}{\bibinfo{person}{Karoliina Isoaho}, \bibinfo{person}{Daria
  Gritsenko}, {and} \bibinfo{person}{Eetu M{\"a}kel{\"a}}.}
  \bibinfo{year}{2021}\natexlab{}.
\newblock \showarticletitle{Topic {{Modeling}} and {{Text Analysis}} for
  {{Qualitative Policy Research}}}.
\newblock \bibinfo{journal}{\emph{Policy Studies Journal}}
  \bibinfo{volume}{49}, \bibinfo{number}{1} (\bibinfo{date}{Feb.}
  \bibinfo{year}{2021}).
\newblock


\bibitem[Jacobs and Wallach(2021)]%
        {jacobs2021measurement}
\bibfield{author}{\bibinfo{person}{Abigail~Z. Jacobs} {and}
  \bibinfo{person}{Hanna Wallach}.} \bibinfo{year}{2021}\natexlab{}.
\newblock \showarticletitle{Measurement and {{Fairness}}}. In
  \bibinfo{booktitle}{\emph{Proceedings of the 2021 {{ACM Conference}} on
  {{Fairness}}, {{Accountability}}, and {{Transparency}}}}.
  \bibinfo{publisher}{{Association for Computing Machinery}}.
\newblock


\bibitem[Jacobs and Tsch{\"o}tschel(2019)]%
        {jacobs2019topic}
\bibfield{author}{\bibinfo{person}{Thomas Jacobs} {and} \bibinfo{person}{Robin
  Tsch{\"o}tschel}.} \bibinfo{year}{2019}\natexlab{}.
\newblock \showarticletitle{Topic Models Meet Discourse Analysis: A
  Quantitative Tool for a Qualitative Approach}.
\newblock \bibinfo{journal}{\emph{International Journal of Social Research
  Methodology}} \bibinfo{volume}{22}, \bibinfo{number}{5}
  (\bibinfo{date}{Sept.} \bibinfo{year}{2019}).
\newblock


\bibitem[Jakobson(1960)]%
        {jakobson1960linguistics}
\bibfield{author}{\bibinfo{person}{Roman Jakobson}.}
  \bibinfo{year}{1960}\natexlab{}.
\newblock \showarticletitle{Linguistics and Poetics}.
\newblock In \bibinfo{booktitle}{\emph{Style in Language}}.
  \bibinfo{publisher}{{MIT Press}}, \bibinfo{address}{{MA}}.
\newblock


\bibitem[Juneau et~al\mbox{.}(2021)]%
        {juneau2021jupyterenabled}
\bibfield{author}{\bibinfo{person}{St{\'e}phanie Juneau}, \bibinfo{person}{Knut
  Olsen}, \bibinfo{person}{Robert Nikutta}, \bibinfo{person}{Alice Jacques},
  {and} \bibinfo{person}{Stephen Bailey}.} \bibinfo{year}{2021}\natexlab{}.
\newblock \showarticletitle{Jupyter-{{Enabled Astrophysical Analysis Using
  Data-Proximate Computing Platforms}}}.
\newblock \bibinfo{journal}{\emph{Computing in Science Engineering}}
  \bibinfo{volume}{23}, \bibinfo{number}{2} (\bibinfo{date}{March}
  \bibinfo{year}{2021}).
\newblock


\bibitem[Kaur et~al\mbox{.}(2020)]%
        {kaur2020interpreting}
\bibfield{author}{\bibinfo{person}{Harmanpreet Kaur}, \bibinfo{person}{Harsha
  Nori}, \bibinfo{person}{Samuel Jenkins}, \bibinfo{person}{Rich Caruana},
  \bibinfo{person}{Hanna Wallach}, {and} \bibinfo{person}{Jennifer~Wortman
  Vaughan}.} \bibinfo{year}{2020}\natexlab{}.
\newblock \showarticletitle{Interpreting {{Interpretability}}: {{Understanding
  Data Scientists}}' {{Use}} of {{Interpretability Tools}} for {{Machine
  Learning}}}.
\newblock \bibinfo{journal}{\emph{Conference on Human Factors in Computing
  Systems - Proceedings}} (\bibinfo{year}{2020}).
\newblock


\bibitem[Kery et~al\mbox{.}(2018)]%
        {kery2018story}
\bibfield{author}{\bibinfo{person}{Mary~Beth Kery}, \bibinfo{person}{Marissa
  Radensky}, \bibinfo{person}{Mahima Arya}, \bibinfo{person}{Bonnie~E. John},
  {and} \bibinfo{person}{Brad~A. Myers}.} \bibinfo{year}{2018}\natexlab{}.
\newblock \showarticletitle{The {{Story}} in the {{Notebook}}: {{Exploratory
  Data Science}} Using a {{Literate Programming Tool}}}. In
  \bibinfo{booktitle}{\emph{Proceedings of the 2018 {{CHI Conference}} on
  {{Human Factors}} in {{Computing Systems}}}}. \bibinfo{publisher}{{ACM}},
  \bibinfo{address}{{Montreal QC Canada}}.
\newblock


\bibitem[Keyes(2018)]%
        {keyes2018misgendering}
\bibfield{author}{\bibinfo{person}{Os Keyes}.} \bibinfo{year}{2018}\natexlab{}.
\newblock \showarticletitle{The Misgendering Machines: {{Trans}}/{{HCI}}
  Implications of Automatic Gender Recognition}.
\newblock \bibinfo{journal}{\emph{Proceedings of the ACM on Human-Computer
  Interaction}} \bibinfo{volume}{2}, \bibinfo{number}{CSCW}
  (\bibinfo{year}{2018}).
\newblock


\bibitem[Kitchin(2014)]%
        {kitchin2014big}
\bibfield{author}{\bibinfo{person}{Rob Kitchin}.}
  \bibinfo{year}{2014}\natexlab{}.
\newblock \showarticletitle{Big {{Data}}, New Epistemologies and Paradigm
  Shifts}.
\newblock \bibinfo{journal}{\emph{Big Data \& Society}} \bibinfo{volume}{1},
  \bibinfo{number}{1} (\bibinfo{date}{April} \bibinfo{year}{2014}).
\newblock


\bibitem[Kluyver et~al\mbox{.}(2016)]%
        {kluyver2016jupyter}
\bibfield{author}{\bibinfo{person}{Thomas Kluyver}, \bibinfo{person}{Benjamin
  {Ragan-Kelley}}, \bibinfo{person}{Fernando P{\'e}rez}, \bibinfo{person}{Brian
  Granger}, \bibinfo{person}{Matthias Bussonnier}, \bibinfo{person}{Jonathan
  Frederic}, \bibinfo{person}{Kyle Kelley}, \bibinfo{person}{Jessica Hamrick},
  \bibinfo{person}{Jason Grout}, \bibinfo{person}{Sylvain Corlay},
  \bibinfo{person}{Paul Ivanov}, \bibinfo{person}{Dami{\'a}n Avila},
  \bibinfo{person}{Safia Abdalla}, \bibinfo{person}{Carol Willing}, {and}
  \bibinfo{person}{{Jupyter development team}}.}
  \bibinfo{year}{2016}\natexlab{}.
\newblock \showarticletitle{Jupyter {{Notebooks}} \textendash{} a Publishing
  Format for Reproducible Computational Workflows}. In
  \bibinfo{booktitle}{\emph{20th {{International Conference}} on {{Electronic
  Publishing}} (01/01/16)}}, \bibfield{editor}{\bibinfo{person}{Fernando
  Loizides} {and} \bibinfo{person}{Birgit Scmidt}} (Eds.).
  \bibinfo{publisher}{{IOS Press}}.
\newblock


\bibitem[Koenecke et~al\mbox{.}(2020)]%
        {koenecke2020racial}
\bibfield{author}{\bibinfo{person}{Allison Koenecke}, \bibinfo{person}{Andrew
  Nam}, \bibinfo{person}{Emily Lake}, \bibinfo{person}{Joe Nudell},
  \bibinfo{person}{Minnie Quartey}, \bibinfo{person}{Zion Mengesha},
  \bibinfo{person}{Connor Toups}, \bibinfo{person}{John~R Rickford},
  \bibinfo{person}{Dan Jurafsky}, {and} \bibinfo{person}{Sharad Goel}.}
  \bibinfo{year}{2020}\natexlab{}.
\newblock \showarticletitle{Racial disparities in automated speech
  recognition}.
\newblock \bibinfo{journal}{\emph{Proceedings of the National Academy of
  Sciences}} \bibinfo{volume}{117}, \bibinfo{number}{14}
  (\bibinfo{year}{2020}), \bibinfo{pages}{7684--7689}.
\newblock


\bibitem[Koenzen et~al\mbox{.}(2020)]%
        {koenzen2020code}
\bibfield{author}{\bibinfo{person}{Andreas~P. Koenzen},
  \bibinfo{person}{Neil~A. Ernst}, {and} \bibinfo{person}{Margaret-Anne~D.
  Storey}.} \bibinfo{year}{2020}\natexlab{}.
\newblock \showarticletitle{Code Duplication and Reuse in {{Jupyter}}
  Notebooks}. In \bibinfo{booktitle}{\emph{2020 {{IEEE Symposium}} on {{Visual
  Languages}} and {{Human-Centric Computing}} ({{VL}}/{{HCC}})}}.
  \bibinfo{publisher}{{IEEE}}.
\newblock


\bibitem[Krafft et~al\mbox{.}(2020)]%
        {krafft2020defining}
\bibfield{author}{\bibinfo{person}{P.~M. Krafft}, \bibinfo{person}{Meg Young},
  \bibinfo{person}{Michael Katell}, \bibinfo{person}{Karen Huang}, {and}
  \bibinfo{person}{Ghislain Bugingo}.} \bibinfo{year}{2020}\natexlab{}.
\newblock \showarticletitle{Defining {{AI}} in {{Policy}} versus {{Practice}}}.
  In \bibinfo{booktitle}{\emph{Proceedings of the {{AAAI}}/{{ACM Conference}}
  on {{AI}}, {{Ethics}}, and {{Society}}}}. \bibinfo{publisher}{{Association
  for Computing Machinery}}.
\newblock


\bibitem[Langenkamp and Yue(2022)]%
        {langenkamp2022how}
\bibfield{author}{\bibinfo{person}{Max Langenkamp} {and}
  \bibinfo{person}{Daniel~N. Yue}.} \bibinfo{year}{2022}\natexlab{}.
\newblock \showarticletitle{How {{Open Source Machine Learning Software Shapes
  AI}}}. In \bibinfo{booktitle}{\emph{Proceedings of the 2022 {{AAAI}}/{{ACM
  Conference}} on {{AI}}, {{Ethics}}, and {{Society}}}}.
  \bibinfo{publisher}{{ACM}}, \bibinfo{address}{{Oxford United Kingdom}}.
\newblock


\bibitem[Larkin(2013)]%
        {larkin2013politics}
\bibfield{author}{\bibinfo{person}{Brian Larkin}.}
  \bibinfo{year}{2013}\natexlab{}.
\newblock \showarticletitle{The Politics and Poetics of Infrastructure}.
\newblock \bibinfo{journal}{\emph{Annual Review of Anthropology}}
  \bibinfo{volume}{42} (\bibinfo{year}{2013}).
\newblock


\bibitem[Larkin(2020)]%
        {larkin2020promising}
\bibfield{author}{\bibinfo{person}{Brian Larkin}.}
  \bibinfo{year}{2020}\natexlab{}.
\newblock \showarticletitle{7. {{Promising Forms}}: {{The Political
  Aesthetics}} of {{Infrastructure}}}.
\newblock In \bibinfo{booktitle}{\emph{The {{Promise}} of {{Infrastructure}}}},
  \bibfield{editor}{\bibinfo{person}{Nikhil Anand}, \bibinfo{person}{Akhil
  Gupta}, {and} \bibinfo{person}{Hannah Appel}} (Eds.).
  \bibinfo{publisher}{{Duke University Press}}.
\newblock


\bibitem[Lee and Singh(2021)]%
        {lee2021risk}
\bibfield{author}{\bibinfo{person}{Michelle Seng~Ah Lee} {and}
  \bibinfo{person}{Jatinder Singh}.} \bibinfo{year}{2021}\natexlab{}.
\newblock \showarticletitle{Risk {{Identification Questionnaire}} for
  {{Detecting Unintended Bias}} in the {{Machine Learning Development
  Lifecycle}}}. In \bibinfo{booktitle}{\emph{Proceedings of the 2021
  {{AAAI}}/{{ACM Conference}} on {{AI}}, {{Ethics}}, and {{Society}}}}.
  \bibinfo{publisher}{{ACM}}, \bibinfo{address}{{Virtual Event USA}}.
\newblock


\bibitem[Leonardi(2012)]%
        {leonardi2012materialitya}
\bibfield{author}{\bibinfo{person}{Paul~M. Leonardi}.}
  \bibinfo{year}{2012}\natexlab{}.
\newblock \bibinfo{booktitle}{\emph{Materiality, {{Sociomateriality}}, and
  {{Socio-Technical Systems}}: {{What Do These Terms Mean}}? {{How}} Are {{They
  Related}}? {{Do We Need Them}}?}}
\newblock \bibinfo{type}{{{SSRN Scholarly Paper}}} ID 2129878.
  \bibinfo{institution}{{Social Science Research Network}},
  \bibinfo{address}{{Rochester, NY}}.
\newblock


\bibitem[Leonardi and Barley(2008)]%
        {leonardi2008materiality}
\bibfield{author}{\bibinfo{person}{Paul~M. Leonardi} {and}
  \bibinfo{person}{Stephen~R. Barley}.} \bibinfo{year}{2008}\natexlab{}.
\newblock \showarticletitle{Materiality and {{Change}}: {{Challenges}} to
  {{Building Better Theory}} about {{Technology}} and {{Organizing}}}.
\newblock \bibinfo{journal}{\emph{Information and Organization}}
  (\bibinfo{year}{2008}).
\newblock


\bibitem[Leonardi et~al\mbox{.}(2012)]%
        {leonardi2012materiality}
\bibfield{author}{\bibinfo{person}{Paul~M Leonardi}, \bibinfo{person}{Bonnie~A
  Nardi}, {and} \bibinfo{person}{Jannis Kallinikos}.}
  \bibinfo{year}{2012}\natexlab{}.
\newblock \bibinfo{booktitle}{\emph{Materiality and Organizing: Social
  Interaction in a Technological World}}.
\newblock \bibinfo{publisher}{{Oxford University Press}},
  \bibinfo{address}{{Oxford}}.
\newblock


\bibitem[Lewis(1997)]%
        {lewis1997reuters}
\bibfield{author}{\bibinfo{person}{David~D. Lewis}.}
  \bibinfo{year}{1997}\natexlab{}.
\newblock \bibinfo{title}{{{UCI Machine Learning Repository}}: {{Reuters-21578
  Text Categorization Collection Data Set}}}.
\newblock
\newblock
\urldef\tempurl%
\url{https://archive.ics.uci.edu/ml/datasets/reuters-21578+text+categorization+collection}
\showURL{%
\tempurl}


\bibitem[Li et~al\mbox{.}(2021)]%
        {li2021algorithmic}
\bibfield{author}{\bibinfo{person}{Lan Li}, \bibinfo{person}{Tina Lassiter},
  \bibinfo{person}{Joohee Oh}, {and} \bibinfo{person}{Min~Kyung Lee}.}
  \bibinfo{year}{2021}\natexlab{}.
\newblock \showarticletitle{Algorithmic {{Hiring}} in {{Practice}}:
  {{Recruiter}} and {{HR Professional}}'s {{Perspectives}} on {{AI Use}} in
  {{Hiring}}}. In \bibinfo{booktitle}{\emph{Proceedings of the 2021
  {{AAAI}}/{{ACM Conference}} on {{AI}}, {{Ethics}}, and {{Society}}}}.
  \bibinfo{publisher}{{Association for Computing Machinery}}.
\newblock


\bibitem[Lin et~al\mbox{.}(2015)]%
        {lin2015microsoft}
\bibfield{author}{\bibinfo{person}{Tsung-Yi Lin}, \bibinfo{person}{Michael
  Maire}, \bibinfo{person}{Serge Belongie}, \bibinfo{person}{Lubomir Bourdev},
  \bibinfo{person}{Ross Girshick}, \bibinfo{person}{James Hays},
  \bibinfo{person}{Pietro Perona}, \bibinfo{person}{Deva Ramanan},
  \bibinfo{person}{C.~Lawrence Zitnick}, {and} \bibinfo{person}{Piotr
  Doll{\'a}r}.} \bibinfo{year}{2015}\natexlab{}.
\newblock \bibinfo{title}{Microsoft {{COCO}}: {{Common Objects}} in
  {{Context}}}.
\newblock
\newblock
\showeprint[arxiv]{arXiv:1405.0312}


\bibitem[Lindstedt(2019)]%
        {lindstedt2019structural}
\bibfield{author}{\bibinfo{person}{Nathan~C. Lindstedt}.}
  \bibinfo{year}{2019}\natexlab{}.
\newblock \showarticletitle{Structural {{Topic Modeling For Social
  Scientists}}: {{A Brief Case Study}} with {{Social Movement Studies
  Literature}}, 2005\textendash 2017}.
\newblock \bibinfo{journal}{\emph{Social Currents}} \bibinfo{volume}{6},
  \bibinfo{number}{4} (\bibinfo{date}{Aug.} \bibinfo{year}{2019}).
\newblock


\bibitem[Mackenzie(2015)]%
        {mackenzie2015production}
\bibfield{author}{\bibinfo{person}{Adrian Mackenzie}.}
  \bibinfo{year}{2015}\natexlab{}.
\newblock \showarticletitle{The Production of Prediction: {{What}} Does Machine
  Learning Want?}
\newblock \bibinfo{journal}{\emph{European Journal of Cultural Studies}}
  \bibinfo{volume}{18}, \bibinfo{number}{4-5} (\bibinfo{year}{2015}).
\newblock


\bibitem[Madaio et~al\mbox{.}(2022)]%
        {madaio2022assessing}
\bibfield{author}{\bibinfo{person}{Michael Madaio}, \bibinfo{person}{Lisa
  Egede}, \bibinfo{person}{Hariharan Subramonyam}, \bibinfo{person}{Jennifer
  Wortman~Vaughan}, {and} \bibinfo{person}{Hanna Wallach}.}
  \bibinfo{year}{2022}\natexlab{}.
\newblock \showarticletitle{Assessing the {{Fairness}} of {{AI Systems}}: {{AI
  Practitioners}}' {{Processes}}, {{Challenges}}, and {{Needs}} for
  {{Support}}}.
\newblock \bibinfo{journal}{\emph{Proceedings of the ACM on Human-Computer
  Interaction}} \bibinfo{volume}{6}, \bibinfo{number}{CSCW1}
  (\bibinfo{date}{March} \bibinfo{year}{2022}).
\newblock


\bibitem[Maier et~al\mbox{.}(2018)]%
        {maier2018applying}
\bibfield{author}{\bibinfo{person}{Daniel Maier}, \bibinfo{person}{A.
  Waldherr}, \bibinfo{person}{P. Miltner}, \bibinfo{person}{G. Wiedemann},
  \bibinfo{person}{A. Niekler}, \bibinfo{person}{A. Keinert},
  \bibinfo{person}{B. Pfetsch}, \bibinfo{person}{G. Heyer}, \bibinfo{person}{U.
  Reber}, \bibinfo{person}{T. H{\"a}ussler}, \bibinfo{person}{H.
  {Schmid-Petri}}, {and} \bibinfo{person}{S. Adam}.}
  \bibinfo{year}{2018}\natexlab{}.
\newblock \showarticletitle{Applying {{LDA Topic Modeling}} in {{Communication
  Research}}: {{Toward}} a {{Valid}} and {{Reliable Methodology}}}.
\newblock \bibinfo{journal}{\emph{Communication Methods and Measures}}
  \bibinfo{volume}{12}, \bibinfo{number}{2-3} (\bibinfo{date}{April}
  \bibinfo{year}{2018}).
\newblock


\bibitem[Mimno et~al\mbox{.}(2011)]%
        {mimno2011optimizing}
\bibfield{author}{\bibinfo{person}{David Mimno}, \bibinfo{person}{Hanna
  Wallach}, \bibinfo{person}{Edmund Talley}, \bibinfo{person}{Miriam Leenders},
  {and} \bibinfo{person}{Andrew McCallum}.} \bibinfo{year}{2011}\natexlab{}.
\newblock \showarticletitle{Optimizing Semantic Coherence in Topic Models}. In
  \bibinfo{booktitle}{\emph{Proceedings of the 2011 Conference on Empirical
  Methods in Natural Language Processing}}.
\newblock


\bibitem[Mohr and Bogdanov(2013)]%
        {mohr2013introduction}
\bibfield{author}{\bibinfo{person}{John~W. Mohr} {and} \bibinfo{person}{Petko
  Bogdanov}.} \bibinfo{year}{2013}\natexlab{}.
\newblock \showarticletitle{Introduction\textemdash{{Topic}} Models: {{What}}
  They Are and Why They Matter}.
\newblock \bibinfo{journal}{\emph{Poetics}} \bibinfo{volume}{41},
  \bibinfo{number}{6} (\bibinfo{date}{Dec.} \bibinfo{year}{2013}).
\newblock


\bibitem[Morley et~al\mbox{.}(2020)]%
        {morley2020what}
\bibfield{author}{\bibinfo{person}{Jessica Morley}, \bibinfo{person}{Luciano
  Floridi}, \bibinfo{person}{Libby Kinsey}, {and} \bibinfo{person}{Anat
  Elhalal}.} \bibinfo{year}{2020}\natexlab{}.
\newblock \showarticletitle{From {{What}} to {{How}}: {{An Initial Review}} of
  {{Publicly Available AI Ethics Tools}}, {{Methods}} and {{Research}} to
  {{Translate Principles}} into {{Practices}}}.
\newblock \bibinfo{journal}{\emph{Science and Engineering Ethics}}
  \bibinfo{volume}{26}, \bibinfo{number}{4} (\bibinfo{date}{Aug.}
  \bibinfo{year}{2020}).
\newblock


\bibitem[Moss(2022)]%
        {moss2022objective}
\bibfield{author}{\bibinfo{person}{Emanuel Moss}.}
  \bibinfo{year}{2022}\natexlab{}.
\newblock \bibinfo{title}{The {{Objective Function}}: {{Science}} and
  {{Society}} in the {{Age}} of {{Machine Intelligence}}}.
\newblock
\newblock
\showeprint[arxiv]{2209.10418}~[cs]


\bibitem[Mueller(2018)]%
        {mueller2018reasons}
\bibfield{author}{\bibinfo{person}{Alexander Mueller}.}
  \bibinfo{year}{2018}\natexlab{}.
\newblock \bibinfo{title}{5 Reasons Why Jupyter Notebooks Suck}.
\newblock
\newblock
\urldef\tempurl%
\url{https://towardsdatascience.com/5-reasons-why-jupyter-notebooks-suck-4dc201e27086}
\showURL{%
\tempurl}


\bibitem[Nasehi et~al\mbox{.}(2012)]%
        {nasehi2012what}
\bibfield{author}{\bibinfo{person}{Seyed~Mehdi Nasehi},
  \bibinfo{person}{Jonathan Sillito}, \bibinfo{person}{Frank Maurer}, {and}
  \bibinfo{person}{Chris Burns}.} \bibinfo{year}{2012}\natexlab{}.
\newblock \showarticletitle{What Makes a Good Code Example?: {{A}} Study of
  Programming {{Q}}\&{{A}} in {{StackOverflow}}}. In
  \bibinfo{booktitle}{\emph{2012 28th {{IEEE International Conference}} on
  {{Software Maintenance}} ({{ICSM}})}}.
\newblock


\bibitem[Newman and Girvan(2004)]%
        {newman2004finding}
\bibfield{author}{\bibinfo{person}{M.~E.~J. Newman} {and} \bibinfo{person}{M.
  Girvan}.} \bibinfo{year}{2004}\natexlab{}.
\newblock \showarticletitle{Finding and Evaluating Community Structure in
  Networks}.
\newblock \bibinfo{journal}{\emph{Physical Review E}} \bibinfo{volume}{69},
  \bibinfo{number}{2} (\bibinfo{date}{Feb.} \bibinfo{year}{2004}).
\newblock


\bibitem[Nikolenko et~al\mbox{.}(2017)]%
        {nikolenko2017topic}
\bibfield{author}{\bibinfo{person}{Sergey~I. Nikolenko},
  \bibinfo{person}{Sergei Koltcov}, {and} \bibinfo{person}{Olessia Koltsova}.}
  \bibinfo{year}{2017}\natexlab{}.
\newblock \showarticletitle{Topic Modelling for Qualitative Studies}.
\newblock \bibinfo{journal}{\emph{Journal of Information Science}}
  \bibinfo{volume}{43}, \bibinfo{number}{1} (\bibinfo{date}{Feb.}
  \bibinfo{year}{2017}).
\newblock


\bibitem[Obermeyer and Mullainathan(2019)]%
        {obermeyer2019dissecting}
\bibfield{author}{\bibinfo{person}{Ziad Obermeyer} {and}
  \bibinfo{person}{Sendhil Mullainathan}.} \bibinfo{year}{2019}\natexlab{}.
\newblock \showarticletitle{Dissecting {{Racial Bias}} in an {{Algorithm That
  Guides Health Decisions}} for 70 {{Million People}}}. In
  \bibinfo{booktitle}{\emph{Proceedings of the {{Conference}} on {{Fairness}},
  {{Accountability}}, and {{Transparency}}}}. \bibinfo{publisher}{{Association
  for Computing Machinery}}.
\newblock


\bibitem[Ophir et~al\mbox{.}(2020)]%
        {ophir2020collaborative}
\bibfield{author}{\bibinfo{person}{Yotam Ophir}, \bibinfo{person}{Dror Walter},
  {and} \bibinfo{person}{Eleanor~R Marchant}.} \bibinfo{year}{2020}\natexlab{}.
\newblock \showarticletitle{A {{Collaborative Way}} of {{Knowing}}: {{Bridging
  Computational Communication Research}} and {{Grounded Theory Ethnography}}}.
\newblock \bibinfo{journal}{\emph{Journal of Communication}}
  \bibinfo{volume}{70}, \bibinfo{number}{3} (\bibinfo{date}{June}
  \bibinfo{year}{2020}).
\newblock


\bibitem[Orr and Davis(2020)]%
        {orr2020attributions}
\bibfield{author}{\bibinfo{person}{Will Orr} {and} \bibinfo{person}{Jenny~L.
  Davis}.} \bibinfo{year}{2020}\natexlab{}.
\newblock \showarticletitle{Attributions of Ethical Responsibility by
  {{Artificial Intelligence}} Practitioners}.
\newblock \bibinfo{journal}{\emph{Information Communication and Society}}
  \bibinfo{volume}{23}, \bibinfo{number}{5} (\bibinfo{year}{2020}).
\newblock


\bibitem[Paris(2021)]%
        {paris2021time}
\bibfield{author}{\bibinfo{person}{Britt~S Paris}.}
  \bibinfo{year}{2021}\natexlab{}.
\newblock \showarticletitle{Time Constructs: {{Design}} Ideology and a Future
  Internet}.
\newblock \bibinfo{journal}{\emph{Time \& Society}} \bibinfo{volume}{30},
  \bibinfo{number}{1} (\bibinfo{date}{Feb.} \bibinfo{year}{2021}).
\newblock
\showISSN{0961-463X, 1461-7463}


\bibitem[Perez and Granger(2007)]%
        {perez2007ipython}
\bibfield{author}{\bibinfo{person}{Fernando Perez} {and}
  \bibinfo{person}{Brian~E. Granger}.} \bibinfo{year}{2007}\natexlab{}.
\newblock \showarticletitle{{{IPython}}: {{A}} System for Interactive
  Scientific Computing}.
\newblock \bibinfo{journal}{\emph{Computing in Science \& Engineering}}
  \bibinfo{volume}{9}, \bibinfo{number}{3} (\bibinfo{year}{2007}).
\newblock


\bibitem[Perkel(2018)]%
        {perkel2018why}
\bibfield{author}{\bibinfo{person}{Jeffrey~M. Perkel}.}
  \bibinfo{year}{2018}\natexlab{}.
\newblock \showarticletitle{Why {{Jupyter}} Is Data Scientists' Computational
  Notebook of Choice}.
\newblock \bibinfo{journal}{\emph{Nature}} \bibinfo{volume}{563},
  \bibinfo{number}{7729} (\bibinfo{date}{Nov.} \bibinfo{year}{2018}).
\newblock


\bibitem[Pimentel et~al\mbox{.}(2019)]%
        {pimentel2019largescale}
\bibfield{author}{\bibinfo{person}{Jo{\~a}o~Felipe Pimentel},
  \bibinfo{person}{Leonardo Murta}, \bibinfo{person}{Vanessa Braganholo}, {and}
  \bibinfo{person}{Juliana Freire}.} \bibinfo{year}{2019}\natexlab{}.
\newblock \showarticletitle{A Large-Scale Study about Quality and
  Reproducibility of Jupyter Notebooks}. In \bibinfo{booktitle}{\emph{2019
  {{IEEE}}/{{ACM}} 16th {{International Conference}} on {{Mining Software
  Repositories}} ({{MSR}})}}. \bibinfo{publisher}{{IEEE}}.
\newblock


\bibitem[Plantin and de~Seta(2019)]%
        {plantin2019wechat}
\bibfield{author}{\bibinfo{person}{Jean~Christophe Plantin} {and}
  \bibinfo{person}{Gabriele de Seta}.} \bibinfo{year}{2019}\natexlab{}.
\newblock \showarticletitle{{{WeChat}} as Infrastructure: The
  Techno-Nationalist Shaping of {{Chinese}} Digital Platforms}.
\newblock \bibinfo{journal}{\emph{Chinese Journal of Communication}}
  \bibinfo{volume}{12}, \bibinfo{number}{3} (\bibinfo{year}{2019}).
\newblock


\bibitem[Plantin et~al\mbox{.}(2018a)]%
        {plantin2018reintegrating}
\bibfield{author}{\bibinfo{person}{Jean-Christophe Plantin},
  \bibinfo{person}{Carl Lagoze}, {and} \bibinfo{person}{Paul~N Edwards}.}
  \bibinfo{year}{2018}\natexlab{a}.
\newblock \showarticletitle{Re-Integrating Scholarly Infrastructure: The
  Ambiguous Role of Data Sharing Platforms}.
\newblock \bibinfo{journal}{\emph{Big Data \& Society}} \bibinfo{volume}{5},
  \bibinfo{number}{1} (\bibinfo{date}{Jan.} \bibinfo{year}{2018}).
\newblock
\showISSN{2053-9517, 2053-9517}


\bibitem[Plantin et~al\mbox{.}(2018b)]%
        {plantin2018infrastructure}
\bibfield{author}{\bibinfo{person}{Jean~Christophe Plantin},
  \bibinfo{person}{Carl Lagoze}, \bibinfo{person}{Paul~N. Edwards}, {and}
  \bibinfo{person}{Christian Sandvig}.} \bibinfo{year}{2018}\natexlab{b}.
\newblock \showarticletitle{Infrastructure Studies Meet Platform Studies in the
  Age of {{Google}} and {{Facebook}}}.
\newblock \bibinfo{journal}{\emph{New Media and Society}} \bibinfo{volume}{20},
  \bibinfo{number}{1} (\bibinfo{year}{2018}).
\newblock


\bibitem[Plantin and Punathambekar(2019)]%
        {plantin2019digital}
\bibfield{author}{\bibinfo{person}{Jean~Christophe Plantin} {and}
  \bibinfo{person}{Aswin Punathambekar}.} \bibinfo{year}{2019}\natexlab{}.
\newblock \showarticletitle{{Digital media infrastructures: pipes, platforms,
  and politics}}.
\newblock \bibinfo{journal}{\emph{Media, Culture and Society}}
  \bibinfo{volume}{41}, \bibinfo{number}{2} (\bibinfo{year}{2019}),
  \bibinfo{pages}{163--174}.
\newblock


\bibitem[Polyzotis et~al\mbox{.}(2017)]%
        {polyzotis2017data}
\bibfield{author}{\bibinfo{person}{Neoklis Polyzotis}, \bibinfo{person}{Sudip
  Roy}, \bibinfo{person}{Steven~Euijong Whang}, {and} \bibinfo{person}{Martin
  Zinkevich}.} \bibinfo{year}{2017}\natexlab{}.
\newblock \showarticletitle{Data Management Challenges in Production Machine
  Learning}.
\newblock \bibinfo{journal}{\emph{Proceedings of the ACM SIGMOD International
  Conference on Management of Data}}  \bibinfo{volume}{Part F1277}
  (\bibinfo{year}{2017}).
\newblock


\bibitem[Polyzotis et~al\mbox{.}(2018)]%
        {polyzotis2018data}
\bibfield{author}{\bibinfo{person}{Neoklis Polyzotis}, \bibinfo{person}{Sudip
  Roy}, \bibinfo{person}{Steven~Euijong Whang}, {and} \bibinfo{person}{Martin
  Zinkevich}.} \bibinfo{year}{2018}\natexlab{}.
\newblock \showarticletitle{Data {{Lifecycle Challenges}} in {{Production
  Machine Learning}}: {{A Survey}}}.
\newblock \bibinfo{journal}{\emph{ACM SIGMOD Record}} \bibinfo{volume}{47},
  \bibinfo{number}{2} (\bibinfo{date}{Dec.} \bibinfo{year}{2018}).
\newblock


\bibitem[Quinn et~al\mbox{.}(2010)]%
        {quinn2010how}
\bibfield{author}{\bibinfo{person}{Kevin~M. Quinn}, \bibinfo{person}{Burt~L.
  Monroe}, \bibinfo{person}{Michael Colaresi}, \bibinfo{person}{Michael~H.
  Crespin}, {and} \bibinfo{person}{Dragomir~R. Radev}.}
  \bibinfo{year}{2010}\natexlab{}.
\newblock \showarticletitle{How to {{Analyze Political Attention}} with
  {{Minimal Assumptions}} and {{Costs}}}.
\newblock \bibinfo{journal}{\emph{American Journal of Political Science}}
  \bibinfo{volume}{54}, \bibinfo{number}{1} (\bibinfo{date}{Jan.}
  \bibinfo{year}{2010}).
\newblock


\bibitem[Rakova et~al\mbox{.}(2021)]%
        {rakova2021where}
\bibfield{author}{\bibinfo{person}{Bogdana Rakova}, \bibinfo{person}{Jingying
  Yang}, \bibinfo{person}{Henriette Cramer}, {and} \bibinfo{person}{Rumman
  Chowdhury}.} \bibinfo{year}{2021}\natexlab{}.
\newblock \showarticletitle{Where {{Responsible AI}} Meets {{Reality}}:
  {{Practitioner Perspectives}} on {{Enablers}} for {{Shifting Organizational
  Practices}}}.
\newblock \bibinfo{journal}{\emph{Proceedings of the ACM on Human-Computer
  Interaction}} \bibinfo{volume}{5}, \bibinfo{number}{CSCW1}
  (\bibinfo{date}{April} \bibinfo{year}{2021}).
\newblock


\bibitem[Randles et~al\mbox{.}(2017)]%
        {randles2017using}
\bibfield{author}{\bibinfo{person}{Bernadette~M. Randles},
  \bibinfo{person}{Irene~V. Pasquetto}, \bibinfo{person}{Milena~S. Golshan},
  {and} \bibinfo{person}{Christine~L. Borgman}.}
  \bibinfo{year}{2017}\natexlab{}.
\newblock \showarticletitle{Using the {{Jupyter}} Notebook as a Tool for Open
  Science: {{An}} Empirical Study}. In \bibinfo{booktitle}{\emph{2017
  {{ACM}}/{{IEEE Joint Conference}} on {{Digital Libraries}} ({{JCDL}})}}.
  \bibinfo{publisher}{{IEEE}}.
\newblock


\bibitem[Redstr{\"o}m(2005)]%
        {redstrom2005technology}
\bibfield{author}{\bibinfo{person}{Johan Redstr{\"o}m}.}
  \bibinfo{year}{2005}\natexlab{}.
\newblock \showarticletitle{On {{Technology}} as {{Material}} in {{Design}}}.
\newblock \bibinfo{journal}{\emph{Design Philosophy Papers}}
  \bibinfo{volume}{3}, \bibinfo{number}{2} (\bibinfo{date}{June}
  \bibinfo{year}{2005}).
\newblock


\bibitem[Roberts et~al\mbox{.}(2016)]%
        {roberts2016model}
\bibfield{author}{\bibinfo{person}{Margaret~E. Roberts},
  \bibinfo{person}{Brandon~M. Stewart}, {and} \bibinfo{person}{Edoardo~M.
  Airoldi}.} \bibinfo{year}{2016}\natexlab{}.
\newblock \showarticletitle{A {{Model}} of {{Text}} for {{Experimentation}} in
  the {{Social Sciences}}}.
\newblock \bibinfo{journal}{\emph{J. Amer. Statist. Assoc.}}
  \bibinfo{volume}{111}, \bibinfo{number}{515} (\bibinfo{date}{July}
  \bibinfo{year}{2016}).
\newblock


\bibitem[Roberts et~al\mbox{.}(2019)]%
        {roberts2019stm}
\bibfield{author}{\bibinfo{person}{Margaret~E. Roberts},
  \bibinfo{person}{Brandon~M. Stewart}, {and} \bibinfo{person}{Dustin
  Tingley}.} \bibinfo{year}{2019}\natexlab{}.
\newblock \showarticletitle{Stm: {{An R Package}} for {{Structural Topic
  Models}}}.
\newblock \bibinfo{journal}{\emph{Journal of Statistical Software}}
  \bibinfo{volume}{91} (\bibinfo{date}{Oct.} \bibinfo{year}{2019}).
\newblock


\bibitem[Roberts et~al\mbox{.}(2013)]%
        {roberts2013structural}
\bibfield{author}{\bibinfo{person}{Margaret~E. Roberts},
  \bibinfo{person}{Brandon~M. Stewart}, \bibinfo{person}{Dustin Tingley}, {and}
  \bibinfo{person}{Edoardo~M. Airoldi}.} \bibinfo{year}{2013}\natexlab{}.
\newblock \showarticletitle{The Structural Topic Model and Applied Social
  Science}. In \bibinfo{booktitle}{\emph{Advances in Neural Information
  Processing Systems Workshop on Topic Models: Computation, Application, and
  Evaluation}}, Vol.~\bibinfo{volume}{4}. \bibinfo{publisher}{{Harrahs and
  Harveys, Lake Tahoe}}.
\newblock


\bibitem[Roberts et~al\mbox{.}(2014)]%
        {roberts2014structural}
\bibfield{author}{\bibinfo{person}{Margaret~E. Roberts},
  \bibinfo{person}{Brandon~M. Stewart}, \bibinfo{person}{Dustin Tingley},
  \bibinfo{person}{Christopher Lucas}, \bibinfo{person}{Jetson {Leder-Luis}},
  \bibinfo{person}{Shana~Kushner Gadarian}, \bibinfo{person}{Bethany
  Albertson}, {and} \bibinfo{person}{David~G. Rand}.}
  \bibinfo{year}{2014}\natexlab{}.
\newblock \showarticletitle{Structural {{Topic Models}} for {{Open-Ended Survey
  Responses}}}.
\newblock \bibinfo{journal}{\emph{American Journal of Political Science}}
  \bibinfo{volume}{58}, \bibinfo{number}{4} (\bibinfo{year}{2014}).
\newblock


\bibitem[Rosen and Shihab(2016)]%
        {rosen2016what}
\bibfield{author}{\bibinfo{person}{Christoffer Rosen} {and}
  \bibinfo{person}{Emad Shihab}.} \bibinfo{year}{2016}\natexlab{}.
\newblock \showarticletitle{What Are Mobile Developers Asking about? {{A}}
  Large Scale Study Using Stack Overflow}.
\newblock \bibinfo{journal}{\emph{Empirical Software Engineering}}
  \bibinfo{volume}{21}, \bibinfo{number}{3} (\bibinfo{date}{June}
  \bibinfo{year}{2016}).
\newblock


\bibitem[Rouse(2007)]%
        {rouse2007practice}
\bibfield{author}{\bibinfo{person}{Joseph Rouse}.}
  \bibinfo{year}{2007}\natexlab{}.
\newblock \showarticletitle{Practice Theory}.
\newblock In \bibinfo{booktitle}{\emph{Philosophy of {{Anthropology}} and
  {{Sociology}}}}. \bibinfo{publisher}{{Elsevier}}.
\newblock


\bibitem[Ruder(2017)]%
        {ruder2016overview}
\bibfield{author}{\bibinfo{person}{Sebastian Ruder}.}
  \bibinfo{year}{2017}\natexlab{}.
\newblock \showarticletitle{An Overview of Gradient Descent Optimization
  Algorithms}.
\newblock  (\bibinfo{year}{2017}).
\newblock
\showeprint[arxiv]{1609.04747}~[cs.LG]


\bibitem[Rule et~al\mbox{.}(2018)]%
        {rule2018exploration}
\bibfield{author}{\bibinfo{person}{Adam Rule}, \bibinfo{person}{Aur{\'e}lien
  Tabard}, {and} \bibinfo{person}{James~D. Hollan}.}
  \bibinfo{year}{2018}\natexlab{}.
\newblock \showarticletitle{Exploration and {{Explanation}} in {{Computational
  Notebooks}}}. In \bibinfo{booktitle}{\emph{Proceedings of the 2018 {{CHI
  Conference}} on {{Human Factors}} in {{Computing Systems}}}}.
  \bibinfo{publisher}{{Association for Computing Machinery}},
  \bibinfo{address}{{New York, NY, USA}}.
\newblock


\bibitem[Ryan et~al\mbox{.}(2021)]%
        {ryan2021research}
\bibfield{author}{\bibinfo{person}{Mark Ryan}, \bibinfo{person}{Josephina
  Antoniou}, \bibinfo{person}{Laurence Brooks}, \bibinfo{person}{Tilimbe Jiya},
  \bibinfo{person}{Kevin Macnish}, {and} \bibinfo{person}{Bernd Stahl}.}
  \bibinfo{year}{2021}\natexlab{}.
\newblock \showarticletitle{Research and {{Practice}} of {{AI Ethics}}: {{A}}
  Case Study Approach Juxtaposing Academic Discourse with Organisational
  Reality}.
\newblock \bibinfo{journal}{\emph{Science and Engineering Ethics}}
  \bibinfo{volume}{27}, \bibinfo{number}{2} (\bibinfo{year}{2021}).
\newblock
\showISSN{1471-5546}


\bibitem[Schaich~Borg(2021)]%
        {schaichborg2021four}
\bibfield{author}{\bibinfo{person}{Jana Schaich~Borg}.}
  \bibinfo{year}{2021}\natexlab{}.
\newblock \showarticletitle{Four Investment Areas for Ethical {{AI}}:
  {{Transdisciplinary}} Opportunities to Close the Publication-to-Practice
  Gap}.
\newblock \bibinfo{journal}{\emph{Big Data \& Society}} \bibinfo{volume}{8},
  \bibinfo{number}{2} (\bibinfo{date}{July} \bibinfo{year}{2021}).
\newblock


\bibitem[Schiff et~al\mbox{.}(2021)]%
        {schiff2021explaining}
\bibfield{author}{\bibinfo{person}{Daniel Schiff}, \bibinfo{person}{Bogdana
  Rakova}, \bibinfo{person}{Aladdin Ayesh}, \bibinfo{person}{Anat Fanti}, {and}
  \bibinfo{person}{Michael Lennon}.} \bibinfo{year}{2021}\natexlab{}.
\newblock \showarticletitle{Explaining the {{Principles}} to {{Practices Gap}}
  in {{AI}}}.
\newblock \bibinfo{journal}{\emph{IEEE Technology and Society Magazine}}
  \bibinfo{volume}{40}, \bibinfo{number}{2} (\bibinfo{date}{June}
  \bibinfo{year}{2021}).
\newblock


\bibitem[Selbst et~al\mbox{.}(2019)]%
        {selbst2019fairness}
\bibfield{author}{\bibinfo{person}{Andrew~D. Selbst}, \bibinfo{person}{Danah
  Boyd}, \bibinfo{person}{Sorelle~A. Friedler}, \bibinfo{person}{Suresh
  Venkatasubramanian}, {and} \bibinfo{person}{Janet Vertesi}.}
  \bibinfo{year}{2019}\natexlab{}.
\newblock \showarticletitle{Fairness and Abstraction in Sociotechnical
  Systems}. In \bibinfo{booktitle}{\emph{Proceedings of the Conference on
  Fairness, Accountability, and Transparency}} (Atlanta, GA, USA)
  \emph{(\bibinfo{series}{FAT* '19})}. \bibinfo{publisher}{Association for
  Computing Machinery}, \bibinfo{address}{New York, NY, USA},
  \bibinfo{pages}{59–68}.
\newblock


\bibitem[Shelby et~al\mbox{.}(2022)]%
        {shelby2022sociotechnical}
\bibfield{author}{\bibinfo{person}{Renee Shelby}, \bibinfo{person}{Shalaleh
  Rismani}, \bibinfo{person}{Kathryn Henne}, \bibinfo{person}{Ajung Moon},
  \bibinfo{person}{Negar Rostamzadeh}, \bibinfo{person}{Paul Nicholas},
  \bibinfo{person}{N'mah Yilla}, \bibinfo{person}{Jess Gallegos},
  \bibinfo{person}{Andrew Smart}, \bibinfo{person}{Emilio Garcia}, {and}
  \bibinfo{person}{Gurleen Virk}.} \bibinfo{year}{2022}\natexlab{}.
\newblock \showarticletitle{Sociotechnical Harms: Scoping a Taxonomy for Harm
  Reduction}.
\newblock  (\bibinfo{date}{Oct.} \bibinfo{year}{2022}).
\newblock
\showeprint[arxiv]{2210.05791}~[cs.HC]


\bibitem[Shove(2003)]%
        {shove2003comfort}
\bibfield{author}{\bibinfo{person}{Elizabeth Shove}.}
  \bibinfo{year}{2003}\natexlab{}.
\newblock \bibinfo{booktitle}{\emph{Comfort, {{Cleanliness}} and
  {{Convenience}} : {{The Social Organization}} of {{Normality}}}}.
\newblock \bibinfo{publisher}{{Berg}}.
\newblock


\bibitem[Shove(2016)]%
        {shove2016matters}
\bibfield{author}{\bibinfo{person}{Elizabeth Shove}.}
  \bibinfo{year}{2016}\natexlab{}.
\newblock \showarticletitle{Matters of Practice}.
\newblock In \bibinfo{booktitle}{\emph{The {{Nexus}} of {{Practices}}}
  (\bibinfo{edition}{1st} ed.)}. \bibinfo{publisher}{{Routledge}}.
\newblock


\bibitem[Shove et~al\mbox{.}(2012)]%
        {shove2012dynamics}
\bibfield{author}{\bibinfo{person}{Elizabeth Shove}, \bibinfo{person}{Mika
  Pantzar}, {and} \bibinfo{person}{Matt Watson}.}
  \bibinfo{year}{2012}\natexlab{}.
\newblock \bibinfo{booktitle}{\emph{The Dynamics of Social Practice: Everyday
  Life and How It Changes}}.
\newblock \bibinfo{publisher}{{SAGE}}, \bibinfo{address}{{Los Angeles}}.
\newblock
\showLCCN{HM831 .S527 2012}


\bibitem[Silvast and Virtanen(2019)]%
        {silvast2019assemblage}
\bibfield{author}{\bibinfo{person}{Antti Silvast} {and}
  \bibinfo{person}{Mikko~J. Virtanen}.} \bibinfo{year}{2019}\natexlab{}.
\newblock \showarticletitle{An Assemblage of Framings and Tamings: Multi-Sited
  Analysis of Infrastructures as a Methodology}.
\newblock \bibinfo{journal}{\emph{Journal of Cultural Economy}}
  \bibinfo{volume}{12}, \bibinfo{number}{6} (\bibinfo{date}{Nov.}
  \bibinfo{year}{2019}).
\newblock
\showISSN{1753-0350, 1753-0369}


\bibitem[Sloane and Zakrzewski(2022)]%
        {sloane2022german}
\bibfield{author}{\bibinfo{person}{Mona Sloane} {and} \bibinfo{person}{Janina
  Zakrzewski}.} \bibinfo{year}{2022}\natexlab{}.
\newblock \showarticletitle{German {{AI Start-Ups}} and ``{{AI Ethics}}'':
  {{Using A Social Practice Lens}} for {{Assessing}} and {{Implementing
  Socio-Technical Innovation}}}. In \bibinfo{booktitle}{\emph{2022 {{ACM
  Conference}} on {{Fairness}}, {{Accountability}}, and {{Transparency}}}}.
  \bibinfo{publisher}{{ACM}}, \bibinfo{address}{{Seoul Republic of Korea}}.
\newblock


\bibitem[Star(1989)]%
        {star1989structure}
\bibfield{author}{\bibinfo{person}{Susan~Leigh Star}.}
  \bibinfo{year}{1989}\natexlab{}.
\newblock \showarticletitle{The {{Structure}} of {{Ill-Structured Solutions}}:
  {{Boundary Objects}} and {{Heterogeneous Distributed Problem Solving}}}.
\newblock In \bibinfo{booktitle}{\emph{Distributed {{Artificial
  Intelligence}}}}. \bibinfo{publisher}{{Elsevier}}.
\newblock


\bibitem[Star(1999)]%
        {star1999ethnography}
\bibfield{author}{\bibinfo{person}{Susan~Leigh Star}.}
  \bibinfo{year}{1999}\natexlab{}.
\newblock \showarticletitle{The Ethnography of Infrastructure}.
\newblock \bibinfo{journal}{\emph{American behavioral scientist}}
  \bibinfo{volume}{43}, \bibinfo{number}{3} (\bibinfo{year}{1999}).
\newblock
\showISSN{0002-7642}


\bibitem[Star and Ruhleder(1996)]%
        {star1996steps}
\bibfield{author}{\bibinfo{person}{Susan~Leigh Star} {and}
  \bibinfo{person}{Karen Ruhleder}.} \bibinfo{year}{1996}\natexlab{}.
\newblock \showarticletitle{Steps {{Toward}} an {{Ecology}} of
  {{Infrastructure}}: {{Design}} and {{Access}} for {{Large Information
  Spaces}}}.
\newblock \bibinfo{journal}{\emph{Information Systems Research}}
  \bibinfo{volume}{7}, \bibinfo{number}{1} (\bibinfo{date}{March}
  \bibinfo{year}{1996}).
\newblock


\bibitem[Syed and Spruit(2017)]%
        {syed2017full}
\bibfield{author}{\bibinfo{person}{Shaheen Syed} {and} \bibinfo{person}{Marco
  Spruit}.} \bibinfo{year}{2017}\natexlab{}.
\newblock \showarticletitle{Full-Text or Abstract? {{Examining}} Topic
  Coherence Scores Using Latent Dirichlet Allocation}. In
  \bibinfo{booktitle}{\emph{2017 {{IEEE International}} Conference on Data
  Science and Advanced Analytics ({{DSAA}})}}. {IEEE}.
\newblock


\bibitem[Szymielewicz et~al\mbox{.}(2020)]%
        {szymielewicz2020where}
\bibfield{author}{\bibinfo{person}{Katarzyna Szymielewicz},
  \bibinfo{person}{Anna Bacciarelli}, \bibinfo{person}{Fanny Hidvegi},
  \bibinfo{person}{Agata Foryciarz}, \bibinfo{person}{Soizic P{\'e}nicaud},
  {and} \bibinfo{person}{Matthias Spielkamp}.} \bibinfo{year}{2020}\natexlab{}.
\newblock \showarticletitle{Where {{Do Algorithmic Accountability}} and
  {{Explainability Frameworks Take Us}} in the {{Real World}}? {{From Theory}}
  to {{Practice}}}. In \bibinfo{booktitle}{\emph{Proceedings of the 2020
  {{Conference}} on {{Fairness}}, {{Accountability}}, and {{Transparency}}}}.
  \bibinfo{publisher}{{Association for Computing Machinery}}.
\newblock


\bibitem[Tahaei et~al\mbox{.}(2020)]%
        {tahaei2020understanding}
\bibfield{author}{\bibinfo{person}{Mohammad Tahaei}, \bibinfo{person}{Kami
  Vaniea}, {and} \bibinfo{person}{Naomi Saphra}.}
  \bibinfo{year}{2020}\natexlab{}.
\newblock \showarticletitle{Understanding Privacy-Related Questions on Stack
  Overflow}. In \bibinfo{booktitle}{\emph{Proceedings of the 2020 {{CHI
  Conference}} on {{Human Factors}} in {{Computing Systems}}}}.
\newblock


\bibitem[Tan(2021)]%
        {tan2021nascent}
\bibfield{author}{\bibinfo{person}{Chiin-Rui Tan}.}
  \bibinfo{year}{2021}\natexlab{}.
\newblock \showarticletitle{The {{Nascent Case}} for {{Adopting Jupyter
  Notebooks}} as a {{Pedagogical Tool}} for {{Interdisciplinary Humanities}},
  {{Social Science}}, and {{Arts Education}}}.
\newblock \bibinfo{journal}{\emph{Computing in Science Engineering}}
  \bibinfo{volume}{23}, \bibinfo{number}{2} (\bibinfo{date}{March}
  \bibinfo{year}{2021}).
\newblock


\bibitem[Treude and Wagner(2019)]%
        {treude2019predicting}
\bibfield{author}{\bibinfo{person}{Christoph Treude} {and}
  \bibinfo{person}{Markus Wagner}.} \bibinfo{year}{2019}\natexlab{}.
\newblock \showarticletitle{Predicting {{Good Configurations}} for {{GitHub}}
  and {{Stack Overflow Topic Models}}}. In \bibinfo{booktitle}{\emph{2019
  {{IEEE}}/{{ACM}} 16th {{International Conference}} on {{Mining Software
  Repositories}} ({{MSR}})}}. \bibinfo{publisher}{{IEEE}},
  \bibinfo{address}{{Montreal, QC, Canada}}.
\newblock


\bibitem[Troeger and Bock(2022)]%
        {troeger2022sociotechnical}
\bibfield{author}{\bibinfo{person}{Jasmin Troeger} {and}
  \bibinfo{person}{Annekatrin Bock}.} \bibinfo{year}{2022}\natexlab{}.
\newblock \showarticletitle{The Sociotechnical Walkthrough \textendash{} a
  Methodological Approach for Platform Studies}.
\newblock \bibinfo{journal}{\emph{Studies in Communication Sciences}}
  \bibinfo{volume}{22}, \bibinfo{number}{1} (\bibinfo{date}{June}
  \bibinfo{year}{2022}).
\newblock
\showISSN{2296-4150, 1424-4896}


\bibitem[Ufford et~al\mbox{.}(2018)]%
        {ufford2018beyond}
\bibfield{author}{\bibinfo{person}{Michelle Ufford}, \bibinfo{person}{M Pacer},
  \bibinfo{person}{Mathew Seal}, {and} \bibinfo{person}{Kyle Kelley}.}
  \bibinfo{year}{2018}\natexlab{}.
\newblock \bibinfo{title}{Beyond {{Interactive}}: {{Notebook Innovation}} at
  {{Netflix}}}.
\newblock
\newblock


\bibitem[Vakkuri et~al\mbox{.}(2020)]%
        {vakkuri2020current}
\bibfield{author}{\bibinfo{person}{Ville Vakkuri},
  \bibinfo{person}{Kai-Kristian Kemell}, \bibinfo{person}{Joni Kultanen}, {and}
  \bibinfo{person}{Pekka Abrahamsson}.} \bibinfo{year}{2020}\natexlab{}.
\newblock \showarticletitle{The {{Current State}} of {{Industrial Practice}} in
  {{Artificial Intelligence Ethics}}}.
\newblock \bibinfo{journal}{\emph{IEEE Software}} \bibinfo{volume}{37},
  \bibinfo{number}{4} (\bibinfo{date}{July} \bibinfo{year}{2020}).
\newblock


\bibitem[Veale and Binns(2017)]%
        {veale2017fairer}
\bibfield{author}{\bibinfo{person}{Michael Veale} {and} \bibinfo{person}{Reuben
  Binns}.} \bibinfo{year}{2017}\natexlab{}.
\newblock \showarticletitle{Fairer Machine Learning in the Real World:
  {{Mitigating}} Discrimination without Collecting Sensitive Data}.
\newblock \bibinfo{journal}{\emph{Big Data \& Society}} \bibinfo{volume}{4},
  \bibinfo{number}{2} (\bibinfo{date}{Dec.} \bibinfo{year}{2017}).
\newblock


\bibitem[Veale et~al\mbox{.}(2018)]%
        {veale2018fairness}
\bibfield{author}{\bibinfo{person}{Michael Veale}, \bibinfo{person}{Max
  Van~Kleek}, {and} \bibinfo{person}{Reuben Binns}.}
  \bibinfo{year}{2018}\natexlab{}.
\newblock \showarticletitle{Fairness and {{Accountability Design Needs}} for
  {{Algorithmic Support}} in {{High-Stakes Public Sector Decision-Making}}}. In
  \bibinfo{booktitle}{\emph{Proceedings of the 2018 {{CHI Conference}} on
  {{Human Factors}} in {{Computing Systems}}}}. \bibinfo{publisher}{{ACM}},
  \bibinfo{address}{{New York, NY, USA}}.
\newblock


\bibitem[Walter and Ophir(2019)]%
        {walter2019news}
\bibfield{author}{\bibinfo{person}{Dror Walter} {and} \bibinfo{person}{Yotam
  Ophir}.} \bibinfo{year}{2019}\natexlab{}.
\newblock \showarticletitle{News {{Frame Analysis}}: {{An Inductive
  Mixed-method Computational Approach}}}.
\newblock \bibinfo{journal}{\emph{Communication Methods and Measures}}
  \bibinfo{volume}{13}, \bibinfo{number}{4} (\bibinfo{date}{Oct.}
  \bibinfo{year}{2019}).
\newblock


\bibitem[Wang et~al\mbox{.}(2019)]%
        {wang2019how}
\bibfield{author}{\bibinfo{person}{April~Yi Wang}, \bibinfo{person}{Anant
  Mittal}, \bibinfo{person}{Christopher Brooks}, {and} \bibinfo{person}{Steve
  Oney}.} \bibinfo{year}{2019}\natexlab{}.
\newblock \showarticletitle{How {{Data Scientists Use Computational Notebooks}}
  for {{Real-Time Collaboration}}}.
\newblock \bibinfo{journal}{\emph{Proceedings of the ACM on Human-Computer
  Interaction}} \bibinfo{volume}{3}, \bibinfo{number}{CSCW}
  (\bibinfo{date}{Nov.} \bibinfo{year}{2019}).
\newblock


\bibitem[Wang et~al\mbox{.}(2020)]%
        {wang2020better}
\bibfield{author}{\bibinfo{person}{Jiawei Wang}, \bibinfo{person}{Li Li}, {and}
  \bibinfo{person}{Andreas Zeller}.} \bibinfo{year}{2020}\natexlab{}.
\newblock \showarticletitle{Better Code, Better Sharing: On the Need of
  Analyzing Jupyter Notebooks}. In \bibinfo{booktitle}{\emph{Proceedings of the
  {{ACM}}/{{IEEE}} 42nd {{International Conference}} on {{Software
  Engineering}}: {{New Ideas}} and {{Emerging Results}}}}.
\newblock


\bibitem[Watson and Shove(2022)]%
        {watson2022how}
\bibfield{author}{\bibinfo{person}{Matt Watson} {and}
  \bibinfo{person}{Elizabeth Shove}.} \bibinfo{year}{2022}\natexlab{}.
\newblock \showarticletitle{How {{Infrastructures}} and {{Practices Shape Each
  Other}}: {{Aggregation}}, {{Integration}} and the {{Introduction}} of {{Gas
  Central Heating}}}.
\newblock \bibinfo{journal}{\emph{Sociological Research Online}}
  (\bibinfo{date}{Jan.} \bibinfo{year}{2022}).
\newblock


\bibitem[Weinberg(2022)]%
        {weinberg2022rethinking}
\bibfield{author}{\bibinfo{person}{Lindsay Weinberg}.}
  \bibinfo{year}{2022}\natexlab{}.
\newblock \showarticletitle{Rethinking {{Fairness}}: {{An Interdisciplinary
  Survey}} of {{Critiques}} of {{Hegemonic ML Fairness Approaches}}}.
\newblock \bibinfo{journal}{\emph{Journal of Artificial Intelligence Research}}
   \bibinfo{volume}{74} (\bibinfo{date}{May} \bibinfo{year}{2022}).
\newblock
\showISSN{1076-9757}


\bibitem[Wesslen(2018)]%
        {wesslen2018computer}
\bibfield{author}{\bibinfo{person}{Ryan Wesslen}.}
  \bibinfo{year}{2018}\natexlab{}.
\newblock \bibinfo{title}{Computer-{{Assisted Text Analysis}} for {{Social
  Science}}: {{Topic Models}} and {{Beyond}}}.
\newblock
\newblock
\showeprint[arxiv]{1803.11045}~[cs]


\bibitem[Winner(1980)]%
        {winner1980artifacts}
\bibfield{author}{\bibinfo{person}{Langdon Winner}.}
  \bibinfo{year}{1980}\natexlab{}.
\newblock \showarticletitle{Do {{Artifacts Have Politics}}?}
\newblock In \bibinfo{booktitle}{\emph{The {{Whale}} and the {{Reactor}} a
  {{Search}} for {{Limits}} in an {{Age}} of {{High Technology}}}}.
  Vol.~\bibinfo{volume}{109}. \bibinfo{publisher}{{The MIT Press}}.
\newblock
\showISSN{00115266}


\bibitem[Yang et~al\mbox{.}(2016)]%
        {yang2016what}
\bibfield{author}{\bibinfo{person}{Xin-Li Yang}, \bibinfo{person}{David Lo},
  \bibinfo{person}{Xin Xia}, \bibinfo{person}{Zhi-Yuan Wan}, {and}
  \bibinfo{person}{Jian-Ling Sun}.} \bibinfo{year}{2016}\natexlab{}.
\newblock \showarticletitle{What {{Security Questions Do Developers Ask}}? {{A
  Large-Scale Study}} of {{Stack Overflow Posts}}}.
\newblock \bibinfo{journal}{\emph{Journal of Computer Science and Technology}}
  \bibinfo{volume}{31}, \bibinfo{number}{5} (\bibinfo{date}{Sept.}
  \bibinfo{year}{2016}).
\newblock


\bibitem[Zhang et~al\mbox{.}(2020)]%
        {zhang2020how}
\bibfield{author}{\bibinfo{person}{Amy~X. Zhang}, \bibinfo{person}{Michael
  Muller}, {and} \bibinfo{person}{Dakuo Wang}.}
  \bibinfo{year}{2020}\natexlab{}.
\newblock \showarticletitle{How Do {{Data Science Workers Collaborate}}?
  {{Roles}}, {{Workflows}}, and {{Tools}}}.
\newblock \bibinfo{journal}{\emph{Proceedings of the ACM on Human-Computer
  Interaction}} \bibinfo{volume}{4}, \bibinfo{number}{CSCW1}
  (\bibinfo{date}{May} \bibinfo{year}{2020}).
\newblock


\end{thebibliography}
